\documentclass[prc,aps,nofootinbib,twocolumn,floatfix]{revtex4}
\usepackage{amsfonts}
\usepackage{amsmath}
\usepackage{amssymb}
\usepackage{epsfig}
\usepackage{slashed}
\usepackage{graphicx}
\usepackage{color}

\newcommand{\be}{\begin{equation}}
\newcommand{\ee}{\end{equation}}
\newcommand{\ba}{\begin{eqnarray}}
\newcommand{\ea}{\end{eqnarray}}
\newcommand{\pa}{\partial}
\newcommand{\unit}{\mathbb{I}}
\newcommand{\nn}{\nonumber}

\begin{document}
\title{Flavor dependence of baryon melting temperature \\ in effective models of QCD}
\author{Juan M. Torres-Rincon, Benjamin Sintes and Joerg Aichelin}
\affiliation{Subatech, UMR 6457, IN2P3/CNRS, Universit\'e de Nantes, \'Ecole de Mines de Nantes, 4 rue Alfred Kastler 44307,
Nantes, France}
\begin{abstract} 
   
   We apply the three-flavor (Polyakov--)Nambu--Jona-Lasinio model to generate baryons as quark-diquark bound
states using many-body techniques at finite temperature. All the baryonic states belonging to the octet and
decuplet flavor representations are generated in the isospin-symmetric case. For each state we extract the
melting temperature at which the baryon may decay into a quark-diquark pair. We seek for an evidence of the
strangeness dependence of the baryon melting temperature as suggested by the statistical thermal models and
supported by lattice-QCD results. A clear and robust signal for this claim is found, pointing to a flavor
dependence of the hadronic deconfinement temperature.
\end{abstract}
\maketitle

\section{Introduction}

   Experiments at the relativistic heavy-ion collider and the large hadron collider (LHC) have shown that a
quark-gluon plasma (QGP) is produced during the first stages of a relativistic heavy-ion collision. The QGP is the phase
of quantum chromodynamics (QCD) at high temperature/density where quarks and gluons are not color-confined into hadrons.
From this QGP phase, the produced fireball undergoes a transition to the hadronic phase at a given hadronization temperature.

At nearly vanishing baryochemical potential the phase transition to the hadronic state is known to be a crossover~\cite{Aoki:2006we}. 
Experimentally it is known that at the so-called chemical freeze-out temperature, which at vanishing chemical potential is
close to the hadronization temperature, the hadrons are in statistical equilibrium. This is the result
of a fit of the hadron abundances in the framework of a statistical model~\cite{BraunMunzinger:1994xr,Andronic:2005yp}.
This fit determines the chemical freeze-out temperature and describes the multiplicity of almost all nonresonant hadrons
with an astonishing precision. After chemical freeze out the hadrons still interact but the chemical composition of the
hadron gas remains (almost) unchanged. Results from high energetic central Pb+Pb collisions at LHC 
show that the freeze-out temperatures extracted by thermal fits~\cite{Preghenella:2011np} are close to the crossover
temperature predicted by lattice-QCD studies~\cite{Bazavov:2011nk}.

   A natural question to ask is whether the freeze-out conditions depend on the hadron species, i.e.,
if the chemical freeze-out temperature depends on flavor. Thermal fits presented in Ref.~\cite{Preghenella:2011np}
show a tension when fitting the different baryonic species with a common freeze-out temperature, suggesting that
the chemical freeze-out temperature for nonstrange baryons is smaller (around 16 MeV) than that for strange
baryons~\cite{Preghenella:2011np,Andronic:2005yp}. In Refs.~\cite{Chatterjee:2013yga,Chatterjee:2014ysa,Chatterjee:2014lfa}
it is shown that thermal fits with two independent freeze-out temperatures (for non-strange and strange hadrons)
provide a better description of the hadronic yields and considerably reduce the $\chi^2$ (per degree of freedom) of the fit. 
This holds for a wide range of collision energies between $\sqrt{s_{NN}}=6.27$ GeV and $\sqrt{s_{NN}}=2.76$ TeV. 
In particular, for Pb+Pb collisions at LHC with $\sqrt{s_{NN}}=2.76$ TeV the difference between the two freeze-out
temperatures is around 15 MeV~\cite{Chatterjee:2013yga}.

   If this is the case, one may suggest that the hadronization temperature also depends on the strangeness
content of the hadron. This idea was brought up quite recently by the authors of Ref.~\cite{Bellwied:2013cta}. In 
this reference, the strangeness dependence of the crossover temperature has been studied with continuum-extrapolated
results of lattice-QCD calculations. The conclusion was that the crossover temperature (measured by the maximum of 
a susceptibility ratio) is about 15 MeV larger for strange hadrons than for those composed by light quarks. This
difference is in surprisingly good accordance with the results from statistical-thermal fits of ALICE abundances, even
if the two physical processes (hadronization and chemical freeze-out) are conceptually distinct. 

    In this paper we study the flavor dependence of the hadronization temperature by using one of the simplest effective
models for strong interactions. The Nambu--Jona-Lasinio (NJL) model is an effective model for low-energy QCD where the gluonic
fields are integrated out and the basic interaction consists of a 4-quark contact vertex. Although the gluon dynamics
is absent in this model, some of the gluonic features can be reproduced by the so-called Polyakov--Nambu--Jona-Lasinio
(PNJL) model. 

    This effective model lacks true confinement. However, hadrons can be thought as dynamically generated
states from multiquark rescattering, thus providing a nonperturbative mechanism for an effective confinement. The
properties of these hadrons (masses and widths) can be obtained by solving the Bethe--Salpeter (BS) equation
(for mesons) and the Fadeev equation (for baryons) with some approximations. Many approaches have been applied
in which meson and baryon properties at zero temperature have been computed within the NJL/PNJL models~\cite{Klevansky:1992qe,Buballa:2003qv,Lehmann,Buck:1992wz,Vogl:1991qt}. 
These models can be extended to finite temperatures and densities. Such an extension allows for calculating
the  ``Mott temperature'', the temperature at which hadrons are not bound anymore, because they can melt into a quark and
a diquark.

  Our aim is to find the Mott temperature for several hadrons within the three-flavor NJL/PNJL models, and extract 
conclusions about its dependence on the strangeness content of the hadrons. In Sec.~\ref{sec:NJL} we introduce
the NJL and PNJL Lagrangians and provide a short remainder on how a meson can be effectively described as a bound state
of a quark-antiquark pair. In Sec.~\ref{sec:diquarks} we use the Bethe-Salpeter equation for two quarks to generate
diquarks and extract their properties as a function of the temperature. In Sec.~\ref{sec:baryon} we apply the Fadeev
equation to generate baryons as bound states of quarks and diquarks. We will present the baryon masses as a function
of temperature up to the melting (or Mott) temperature. We consider all baryons belonging to the octet and decuplet
flavor representations in the isospin limit (up and down quarks with equal masses). Finally, in Sec.~\ref{sec:conclusions} we present our conclusions,
summary and outlook. We also include six appendices with technical details, so keep this paper as much self-contained as
possible: In App.~\ref{app:fierz} we detail the Fierz transformation for the NJL Lagrangian and discuss different versions
found in the literature. Next, in App.~\ref{app:AB} we shortly define the thermal functions needed in the calculation of the 
quark condensate and the meson/diquark polarization function. Appendix~\ref{app:bs} is devoted to the simplification
of the Bethe-Salpeter equation for diquarks and provide the flavor matrices for the different diquark sectors. In
App.~\ref{app:fadeev} we reduce the Fadeev equation for baryons to a tractable form using the ``static approximation''.
In App.~\ref{app:projections} we present the projectors on the physical baryon states belonging to the octet and decuplet
flavor representations. Finally, App.~\ref{app:barpol} is devoted
to the simplification of the quark-diquark polarization function, providing useful expressions.

\section{\label{sec:NJL} Nambu--Jona-Lasinio model for quarks}

\subsection{Effective Lagrangian}

   The NJL model~\cite{Nambu:1961tp,Klevansky:1992qe,Hatsuda:1994pi,Buballa:2003qv} describes the
low-energy interactions of quarks by a four-fermion contact vertex. Being an effective model of QCD it
respects the symmetries of the underlying theory, in particular the $U_V(1) \otimes SU_V(N_f) \otimes SU_A(N_f)$ global
symmetries of the massless QCD Lagrangian. The $U_V(1)$ symmetry leads to the baryon number conservation, while the chiral
symmetry $SU_V(N_f) \otimes SU_A(N_f)$ is spontaneously broken down to $SU_V(N_f)$ at low temperatures. The $U_A(1)$
symmetry is broken by the axial anomaly.

   The form of the NJL Lagrangian is motivated by QCD. Consider the quark-antiquark current-current scattering 
in the $t$ channel, mediated by a dressed gluon. The gluon propagator gives a factor $1/(t-m_g^2)$, with $m_g$ being the dressed gluon mass. In the limit of  low momentum transfer one can
neglect $t$ in comparison with the gluon mass, thus producing an effective contact interaction among quarks.
In this way, the gauge fields disappear from the theory and all what remains is an effective
coupling $g$ between quarks, related to the strong coupling constant and to the gluon mass.

   For three flavors, the resulting effective interaction Lagrangian reads
\be \label{eq:lagNJL} {\cal L} = - g \sum_{a'} \sum_{ij}  (\bar{\psi}_i \ \gamma^\mu T^{a'} \psi_i) \ (\bar{\psi}_j \ \gamma_\mu T^{a'} \psi_j) \ , \ee
where $i,j=1,...,N_f=3$ are flavor indices and $a'=1,...,N_c^2-1$ are color indices with $T^{a'}$ being the color generators, which for $N_c=3$ they are represented by the Gell-Mann matrices,
\be T^{a'} = \lambda^{a'} \ ,  \ee
with the standard normalization 
\be \textrm{tr}_c \  (T^{a'} T^{b'}) = 2\delta^{a'b'}  \ , \ee
where $\textrm{tr}_c$ denote the trace in color space.

   After performing a Fierz transformation~\cite{Klevansky:1992qe,Buballa:2003qv} this Lagrangian can be reexpressed
in a convenient way to describe the $qq$, $\bar{q}\bar{q}$ and $\bar{q}q$ scattering. Using the Fierz transformation described 
in App.~\ref{app:fierz} we first obtain the $\bar{q} q$ sector of the effective theory, which will allow us to describe
mesons. For instance, the pseudoscalar sector of the interacting Lagrangian (\ref{eq:Lex}) reads (all repeated indices are to be summed)
\be \label{eq:fierz1} {\cal L}_{\bar{q}q} = G \ (\bar{\psi}_i \ i\gamma_5 \ \tau^a_{ij} \ \psi_j) 
(\bar{\psi}_k \ i\gamma_5 \ \tau^a_{kl} \ \psi_l) \ , \ee
where $a=1,...,N_f^2-1$ and $G$ is a coupling constant, proportional to the original $g$ in Eq.~(\ref{eq:lagNJL}). In this work, we
will take $G$ as a free parameter to be fixed by comparing our results with the experimental hadron spectrum.
The flavor generators $\tau^a$ follow the normalization
\be \textrm{ tr } (\tau^a \tau^b) = 2 \delta^{ab} \ . \ee
For $N_f=3$ they can be represented by the Gell-Mann matrices.

   The axial anomaly is responsible for the $U_A(1)$ breaking and gives rise to the observed $\eta-\eta'$ mass splitting. To
account for this effect in the our model, we complement the NJL $\bar{q}q$-Lagrangian in Eq.~(\ref{eq:fierz1}) with the 't Hooft Lagrangian:
\be {\cal L}_{H} = - H \det_{ij} \left[ \bar{\psi}_i \ ( \unit - \gamma_5 ) \psi_j \right] - H \det_{ij} \left[ \bar{\psi}_i \ ( \unit + \gamma_5 ) \psi_j \right] \ , \ee 
where $H$ is an additional unknown coupling and $\unit$ is the identity matrix in Dirac space. For $N_f=3$ this Lagrangian 
represents a six-point fermion interaction, which is effectively projected onto a
four-fermion interaction by using the mean-field approximation~\cite{Klevansky:1992qe,Buballa:2003qv}. Using the same approximation,
the quark masses obey the gap equation 
\be \label{eq:gap} m_i = m_{i0} - 4 G \langle \bar{\psi}_i \psi_i \rangle + 2 H \langle \bar{\psi}_j \psi_j \rangle\langle \bar{\psi}_k \psi_k \rangle \ ,
\quad j,k\neq i; j\neq k \ee
with $m_{i0}$ being the bare quark mass for flavor $i$, and the quark condensate defined as
\be \label{eq:conden} \langle \bar{\psi}_i \psi_i \rangle = -i N_c \textrm{ tr}_\gamma \int \frac{d^4k}{(2\pi)^4} S_i (k) \ , \ee
with the trace acting in Dirac space. We represent by $S_i$ the dressed quark propagator,
\be S_i(k) = \frac{1}{\slashed{k}-m_i} \ . \ee
The final expression for the quark condensate in the NJL model is shown in App.~\ref{app:AB}.

\subsection{Medium effects}
To calculate the hadron properties at  finite temperature, we use the imaginary time formalism with the prescription
\be \int \frac{d^4 k}{(2\pi)^4} \rightarrow i T \sum_{n \in \mathbb{Z}} \int \frac{d^3 k}{(2\pi)^3} \ , \ee
with $T$ the temperature and $k^0 \rightarrow i \omega_n$ the fermionic Matsubara frequencies $i\omega_n =i \pi T (2n+1)$.

To account for the finite baryonic density we can introduce a quark chemical potential by adding to
the Lagrangian the term
\be {\cal L}_\mu = \sum_{ij} \bar{\psi}_i \ \mu_{ij} \gamma_0 \ \psi_j \ , \ee
where $\mu_{ij} = \textrm{ diag } (\mu_u,\mu_d,\mu_s)$ contains the quark chemical potentials (which can be 
alternatively expressed in terms of the baryon, charge, and strangeness chemical potentials, $\mu_B,\mu_Q,\mu_S$).
In this work we will restrict ourselves to a vanishing chemical potential $\mu_u=\mu_d=\mu_s=0$.

\subsection{Polyakov--NJL model}

   In the NJL Lagrangian, the gluon fields have been integrated out of the fundamental theory. However, one can
still introduce a source of gluonic effects through the Polyakov line
\be L ({\bf x})= {\cal P} \exp \left( i \int_0^\beta d\tau A_4(\tau,{\bf x}) \right) \ , \ee
where $\beta=1/T$, ${\cal P}$ is the path-ordering operator and $A_4=iA^0$, the temporal component 
of the gluon field in Euclidean space (with $A^\mu= g_s A_a^\mu T_a$).
The order parameter of the deconfinement phase transition (in the absence of quarks) is chosen to be the 
Polyakov loop, $\Phi$, which is the thermal expectation value
\be \Phi = \frac{1}{N_c} \textrm{ tr}_c \langle L \rangle \ , \ee
where the trace in taken in color space.
  To account for deconfinement effects via the Polyakov loop, an effective potential ${\cal U} (\Phi,\bar{\Phi},T)$,
is added to the effective NJL Lagrangian ${\cal L} \rightarrow {\cal L} - {\cal U}$. ${\cal U}$ is a function of the Polyakov loop and its complex conjugate, which
are taken to be homogeneous fields. The form of the effective potential is inspired by the $\mathbb{Z}_3$ center symmetry~\cite{Ratti:2005jh}
\be \frac{{\cal U} (T,\Phi,\bar{\Phi})}{T^4} = - \frac{b_2(T)}{2} \bar{\Phi} \Phi - \frac{b_3}{6}\left( \Phi^3 + 
\bar{\Phi}^3 \right) + \frac{b_4}{4} \left(  \bar{\Phi} \Phi \right)^2 \ , \ee
with
\be b_2(T) = a_0 + a_1 \frac{T_0}{T} + a_2  \left( \frac{T_0}{T} \right)^2 + a_3  \left( \frac{T_0}{T} \right)^3 \ . \ee

The parameters $a_0,a_1,a_2,a_3,b_3,b_4$ and $T_0$ are fitted from the pure-gauge lattice-QCD equation of state at
zero chemical potential~\cite{Ratti:2005jh}. The numerical values of our parameters are given in Table~\ref{tab:param}.
Following the reasoning of Ref.~\cite{Schaefer:2007pw} we have considered the running of $T_0$ with the number of flavors.
As a consequence, we have modified the original parameter $T_0=270$ MeV for the Yang-Mills case ($N_f=0$) to a value
of $T_0=190$ MeV for our case ($N_f=2+1$). 

  This model is called the Polyakov--Nambu--Jona-Lasinio model and has been widely used in similar studies
as ours, e.g., for QCD thermodynamics~\cite{Ratti:2005jh} or generation of bound states~\cite{Hansen:2006ee,Goessens,Blanquier:2011zz}.

  The PNJL grand-canonical potential reads
\ba & & \Omega_{PNJL} (\Phi,\bar{\Phi},m_i,T) = {\cal U}(T,\Phi,\bar{\Phi}) + 2G \sum_i \langle \bar{\psi}_i \psi_i \rangle^2 \nn \\
& -& 4H \prod_i  \langle \bar{\psi}_i \psi_i \rangle  - 2N_c \sum_i \int \frac{d^3 k}{(2\pi)^3} E_i \nn \\
&-& 2 T \sum_i \int \frac{d^3 k}{(2\pi)^3} \left[ \textrm{ tr}_c \log \left(1+L e^{-E_i/T} \right) \right. \nn \\
&+& \left. \textrm{ tr}_c \log \left(1+L^\dag e^{-E_i/T} \right)  \right] \ , 
 \ea
 with $E_i=\sqrt{k^2 + m_i^2}$. Using the mean-field approximation one has~\cite{Hansen:2006ee}
\begin{widetext}
\ba 
 \textrm{ tr}_c \log \left(1+L e^{-E_i/T} \right) &=& \log \left( 1 + 3(\Phi + \bar{\Phi} e^{-E_i/T}) e^{-E_i/T}
+e^{-3E_i/T} \right) \ , \\
 \textrm{ tr}_c \log \left(1+L^\dag e^{-E_i/T} \right) &=& \log \left( 1 + 3(\bar{\Phi} + \Phi e^{-E_i/T}) e^{-E_i/T}
+e^{-3E_i/T} \right) \ . 
\ea
\end{widetext}
We minimize the grand-canonical potential with respect to the order parameters: $\Phi,\bar{\Phi},\langle \bar{\psi}_i \psi_i \rangle$,
\be \label{eq:min} \frac{\pa \Omega_{PNJL}}{\pa \Phi} =0 \ , \quad \frac{\pa \Omega_{PNJL}}{\pa \bar{\Phi}} =0 \ , \quad  
 \frac{\pa \Omega_{PNJL}}{\pa \langle {\bar \psi}_i \psi_i \rangle} =0 \ . \ee

The last equation provides the expression for the quark condensate in the PNJL model,
\be \label{eq:condenpnjl} \langle \bar{\psi}_i \psi_i \rangle = - 2N_c \int  \frac{d^3 k}{(2\pi)^3} \frac{m_i}{E_i} \left[1 - f_\Phi^+(E_i)
- f_\Phi^- (E_i)\right]  \ , \ee
with
\ba f_\Phi^+ (E_i) &=& \frac{ ( \Phi + 2 \bar{\Phi} e^{- E_i/T} ) e^{- E_i/T} + e^{-3  E_i/T}}{1+
 3( \Phi + \bar{\Phi} e^{- E_i/T}) e^{- E_i/T } + e^{-3 E_i/T}} \ , \\
f_\Phi^- (E_i) & =& \frac{ ( \bar{\Phi} + 2 \Phi e^{- E_i/T} ) e^{- E_i/T } + e^{-3 E_i/T}}{1+ 
3( \bar{\Phi} + \Phi e^{- E_i/T}) e^{- E_i/T } + e^{-3  E_i /T}} \  .  \ea

For $N_f=3$, the gap relation (\ref{eq:gap}) and the first two equations in (\ref{eq:min}) form a
system of five coupled equations. The system needs to be solved numerically to obtain the
value of the Polyakov loop (and its conjugate) and the quark masses. In the isospin limit, two equations
are degenerate giving $m_u=m_d$ (and $\langle {\bar \psi}_u \psi_u \rangle=\langle \bar{\psi}_d \psi_d \rangle$).
In addition, at vanishing chemical potential one has $\Phi=\bar{\Phi}$, which is evident from our equations.

\begin{figure*}[htp]
\begin{center}
\includegraphics[scale=0.4]{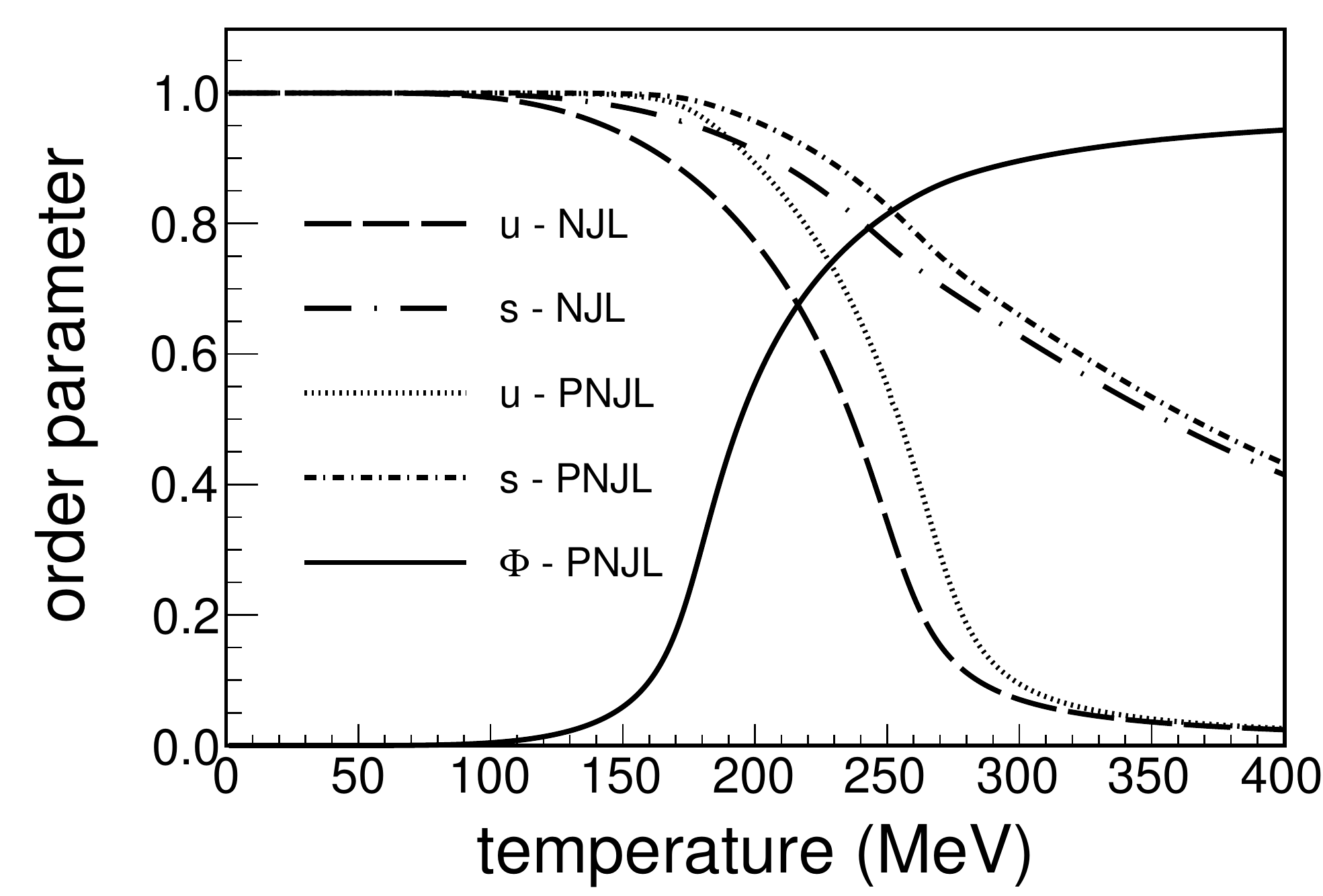}
\includegraphics[scale=0.4]{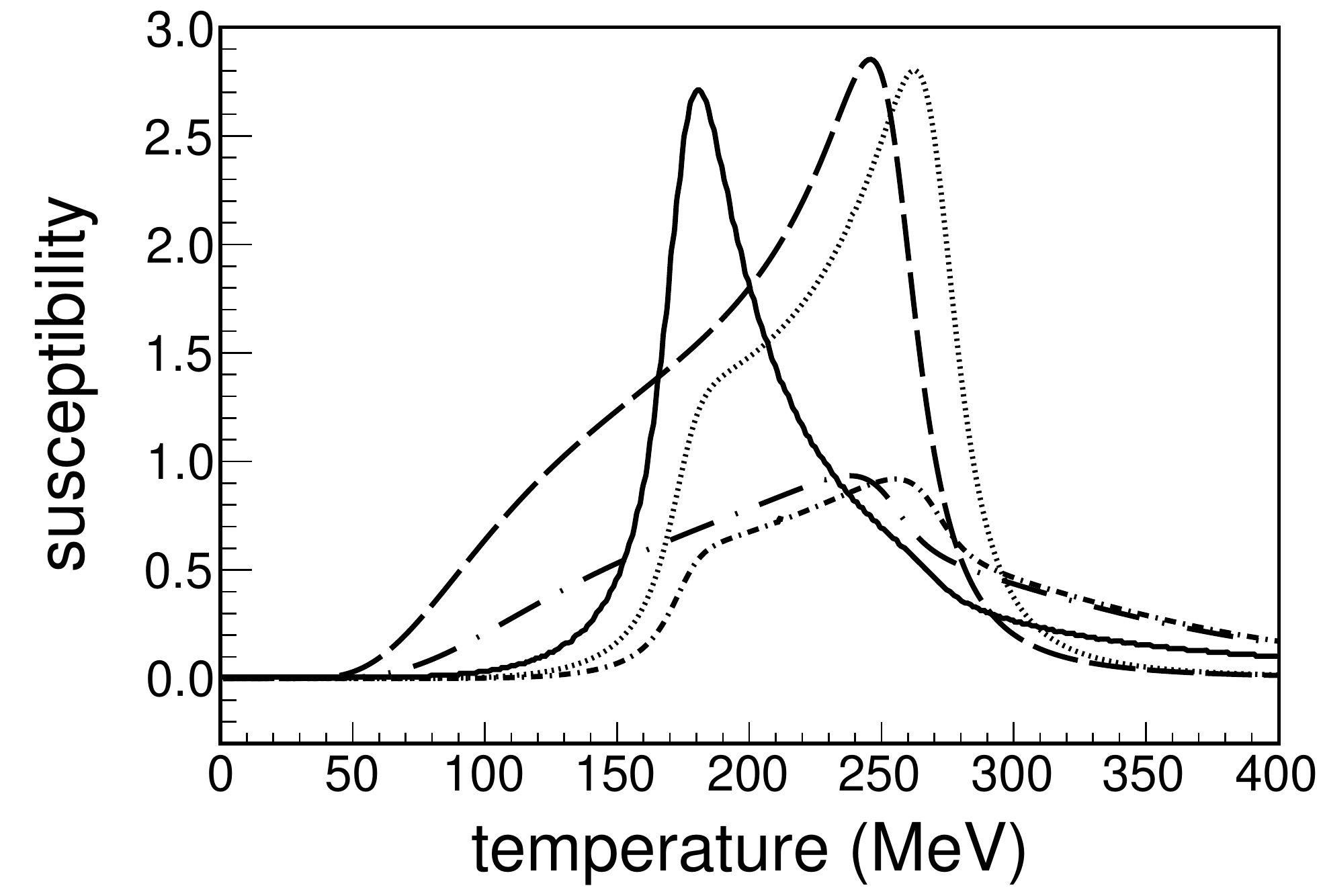}
\caption{\label{fig:transition} Left panel: light and strange quark condensates (for both NJL and PNJL models)
and Polyakov loop as a function of the temperature. In the legend, $u$ and $s$ stand for $\langle {\bar \psi}_u \psi_u \rangle (T)
/\langle {\bar \psi}_u \psi_u \rangle (T=0)$ and  $\langle {\bar \psi}_s \psi_s \rangle (T)
/\langle {\bar \psi}_s \psi_s \rangle (T=0)$, respectively. Right panel: Chiral and deconfinement susceptibilities as a function of
temperature. Their maxima show the chiral and deconfinement transition temperatures. In this panel, the labels $u, d$, and $\Phi$
stand, respectively, for $\chi_{\langle {\bar \psi}_u \psi_u \rangle} (T), \chi_{\langle {\bar \psi}_s \psi_s \rangle} (T)
$ and $\chi_\Phi (T)$.}
\end{center} 
\end{figure*}

In the left panel of Fig.~\ref{fig:transition} we show the temperature dependence of the light and strange quark
condensates, $\langle {\bar \psi}_u \psi_u \rangle$ and $\langle {\bar \psi}_s \psi_s \rangle$ for both 
NJL and PNJL models. In addition, we also show the Polyakov loop as a function of $T$. These quantities serve as
order parameters of the chiral and deconfinement phase transitions. The transition temperature can be defined
as the point at which the susceptibility (derivative of the order parameter) peaks. 
 In the right panel of Fig.~\ref{fig:transition} we plot the dimensionless chiral susceptibilities defined as
\be \chi_{\langle {\bar \psi}_u \psi_u \rangle} \equiv \frac{1}{T^2} \frac{d \langle {\bar \psi}_u \psi_u \rangle}{dT} 
\ , \quad \chi_{\langle {\bar \psi}_s \psi_s \rangle} \equiv \frac{1}{T^2} \frac{d \langle {\bar \psi}_s \psi_s \rangle}{dT} \ . \ee
The maximum of the susceptibility will indicate the chiral transition temperature. They read $T_{\langle {\bar \psi}_u \psi_u \rangle} = 246/262$ MeV (NJL/PNJL)
and $T_{\langle {\bar \psi}_s \psi_s \rangle}=238/255$ MeV (NJL/PNJL). We also show the deconfinement susceptibility defined as
\be \chi_\Phi \equiv T \frac{d\Phi}{dT} \ , \ee
whose maximum gives the approximate position of the deconfinement transition temperature $T_\Phi=181$ MeV. 

\subsection{Mesons as bound states of $\bar{q}q$}      
   To obtain the meson propagator one must solve the BS equation for the quark-antiquark scattering amplitude
$i + \bar{j} \rightarrow m + \bar{n}$ (latin subindices will denote quark flavor and barred indices antiquark
flavor) in the random-phase approximation (RPA):
\begin{widetext}
\be T^{ab}_{i \bar{j},m \bar{n}} (p^2) = {\cal K}^{ab}_{i \bar{j},m \bar{n}} + i \int \frac{d^4 k}{(2 \pi)^4}
{\cal K}^{ac}_{i \bar{j}, p \bar{q}} \  S_{p} \left( k+ \frac{p}{2} \right) \ S_{ \bar{q}}  \left( k-\frac{p}{2} \right)
\ T^{cb}_{p \bar{q},m \bar{n}} (p^2) \ , \ee
\end{widetext}
where $a,b$ denotes the meson flavor channel. The kernel ${\cal K}$ reads
\be {\cal K}^{ab}_{i\bar{j},m\bar{n}} = \Omega^a_{i \bar{j}} \ 2 K^{ab} \ \bar{\Omega}^b_{\bar{n}m} \ , \ee
with the vertex factors containing color, flavor and spin matrices
\be \Omega^a_{i \bar{j}}= \left( \unit_{\textrm{color}} \otimes \tau_{i \bar{j}}^a \otimes \Gamma \right)\ , \ee
as well as a combinatorial factor of 2. The Dirac structure --whose indices we have omitted in our BS equation-- 
can be chosen to be $\Gamma=\{1,i\gamma_5, \gamma^\mu, \gamma_5 \gamma^\mu \}$ for scalar, pseudoscalar, vector, and axial-vector mesons, respectively.

The addition of the 't Hooft term to the NJL Lagrangian breaks flavor symmetry at the mean-field level of
the coupling constants. The resulting flavor-dependent couplings in the $\bar{q}q$ sector $K^{ab}$ are
combinations of the $G$ and $H$ couplings~\cite{Klevansky:1992qe}. In the pseudoscalar sector, the nonzero
couplings read~\cite{Klevansky:1992qe,Rehberg:1995kh} 
\ba K^{00} &=& G + \frac{H}{3} \left( \langle {\bar \psi_u} \psi_u \rangle+\langle {\bar \psi_d} \psi_d \rangle +\langle {\bar \psi_s} \psi_s \rangle \right) \ , \\
 K^{11} &=& K^{22} = K^{33} = G - \frac{H}{2} \langle {\bar \psi_s} \psi_s \rangle \ , \\
K^{44} &=& K^{55} = G - \frac{H}{2} \langle {\bar \psi_d} \psi_d \rangle \ , \\ 
K^{66} &=& K^{77}= G - \frac{H}{2} \langle {\bar \psi_u} \psi_u \rangle \ , \\
K^{88} &=&  G - \frac{H}{6} \left(  2 \langle {\bar \psi_u} \psi_u \rangle +2 \langle {\bar \psi_d} \psi_d \rangle- \langle {\bar \psi_s} \psi_s \rangle \right) \ , \\
K^{03} &=&  K^{30} =  \frac{H}{2\sqrt{6}} \left(  \langle {\bar \psi_u} \psi_u \rangle - \langle {\bar \psi_d} \psi_d \rangle \right) \ , \\
K^{08} &=&  K^{80} =  \frac{-H}{2\sqrt{6}} \left(  \langle {\bar \psi_u} \psi_u \rangle + \langle {\bar \psi_d} \psi_d \rangle- 2  \langle {\bar \psi_s} \psi_s \rangle \right) \ , \\
K^{38} &=&  K^{83} =  -\frac{H}{2\sqrt{3}} \left(  \langle {\bar \psi_u} \psi_u \rangle - \langle {\bar \psi_d} \psi_d \rangle \right) \ . \ea
Notice that they are diagonal in flavor space except for the (0-3-8) subsystem. These non-diagonal couplings will eventually 
bring a $\pi^0-\eta^0-\eta^8$ mixing~\cite{Klevansky:1992qe}, which should be solved in the coupled-channel basis.
In the isospin limit ($m_u=m_d$) the $\pi^0$ is decoupled from the system, but mixing is still present 
in the $\eta^0-\eta^8$ subspace. In a diagonal basis, this fact accounts for the $\eta-\eta'$ mixing, providing
the mass splitting between these two states. Note that in the absence of the 't Hooft term, the mixing disappears.

Introducing the function $t^{ab} (p^2)$
\be T_{i \bar{j},m \bar{n}}^{ab} (p^2) = \Omega^a_{i \bar{j}} \ t^{ab} (p^2) \bar{\Omega}^b_{\bar{n}m} \ , \ee
the solution of the BS equation is a matrix in flavor space
\be \label{eq:meson} t^{ab} (p^2) = \left[ \frac{2K}{1- 2 K \Pi (p^2)} \right]^{ab} \ , \ee
where the polarization function $\Pi^{ab} (p^2)$ is defined as 
\be \label{eq:polmeson} \Pi^{ab} (p^2) = i \int \frac{d^4 k}{(2\pi)^4} \textrm{ tr}_{\gamma}
 \left[\bar{\Omega}^a_{\bar{j}i} S_i \left(k+ \frac{p}{2} \right) \Omega^b_{i\bar{j}} S_{\bar j} \left(k - \frac{p}{2} \right)  \right] \ . \ee
In App.~\ref{app:AB} we provide a simplified expression for this function at finite temperature.

The poles of $t^{ab} (p^2)$ represent the mesonic states, which are bound states of the $\bar{q}q$ scattering.
One can perform a Taylor expansion of the function $t^{-1,ab} (p^2)$ around the pole $p^2=m_M^2$,
\begin{widetext}
\be t^{-1,ab} (p^2)  =  t^{-1,ab} (m_M^2) + \left. \frac{\pa t^{-1,ab} (p^2)}{\pa p^2} \right|_{p^2=m_M^2} (p^2-m_M^2) + \cdots 
\simeq - \frac{1}{2m_M} \left. \frac{\pa \Pi^{ab} (p^2)}{\pa p} \right|_{p^2=m_M^2} (p^2-m_M^2) \ , 
\ee
\end{widetext}
where we have used that $t^{-1,ab} (m_M^2)=0$ at the pole position. 
Defining the effective coupling
\be g^2_{M\rightarrow \bar{q}q} \equiv \frac{2m_M}{ \left. \frac{\pa \Pi^{ab} (p^2)}{\pa p} \right|_{p^2=m_M^2} } \ , \ee
we show that $t^{ab} (p^2)$ can be identified with the meson propagator
\be t^{ab} (p^2)=\frac{-g^2_{M\rightarrow \bar{q}q}}{p^2-m_M^2} \ . \ee 
Therefore, the equation,
\be \label{eq:mesonmass} 1- 2 K^{ab} \Pi^{ab} (p^2=m_M^2) = 0 \ , \ee
gives the meson mass $m_M$ in the appropriate flavor channel (and spin channel by selecting $\Gamma$).

If the generated state has a $m_M$ larger than the sum of quarks masses, it is possible for this meson to decay
into a quark-antiquark pair. In this case the polarization function~(\ref{eq:polmeson}) becomes complex and the pole acquires an
imaginary part. Considering the variable $p^2$ as complex one can identify the mass and the decay width
with the real and imaginary parts of the pole position. In this way one can obtain the meson masses
and decay widths as a function of temperature and/or chemical potential.
A detailed discussion about this procedure is provided at the end of App.~\ref{app:AB}.

The parameters we use in this work are partially based on the findings of Ref.~\cite{Mu:2012zz}.
For the NJL model in the isospin limit we have seven parameters. The extension to the PNJL model
introduces another seven parameters (fixed from the thermodynamics of pure-gauge QCD in the lattice~\cite{Ratti:2005jh}).
All of them are summarized in Table~\ref{tab:param}.
\begin{widetext}
\begin{table*}
\begin{center}
\begin{tabular}{|c|c|c|c|c|c|c|c|}
\hline
Parameter & $m_{q0}$ & $m_{s0}$ & $\Lambda$  &$G$ & $H$ & $G_{DIQ}$ & $G_{DIQ,V}$  \\
\hline
Value & 5.5 MeV & 134 MeV & 569 MeV& $2.3/\Lambda^2$ & $11/\Lambda^5$& $1.56 \ G$& $-0.639 \ G_{DIQ}$ \\
\hline
\hline
Parameter & $a_0$ & $a_1$ & $a_2$ & $a_3$ & $b_3$ & $b_4$ & $T_0$ \\ 
\hline
Value  & 6.75 & -1.95 & 2.625 & -7.44 & 0.75 & 7.5 & 190 MeV \\ 
\hline
\end{tabular}
\caption{\label{tab:param} Parameters of the NJL and PNJL model used in this study. In the isospin limit
we have $m_{q0}=m_{u0}=m_{d0}$.}
\end{center}
\end{table*}
\end{widetext}

  Using the parameter set in Table~\ref{tab:param} we obtain at $T=0$: the light-quark
condensate $\langle \bar{\psi}_u \psi_u\rangle=-(241.3$ MeV$)^3$, the pion decay constant $f_\pi=92.2$ MeV,
the pion mass $m_\pi=134.8$ MeV, the kaon mass $m_K=492.1$ MeV, the $\eta-\eta'$ mass splitting of $475.5$ MeV, the
proton mass $932.0$ MeV and the $\Delta$ baryon mass $1221.4$ MeV.

\begin{figure*}[htp]
\begin{center}
\includegraphics[scale=0.4]{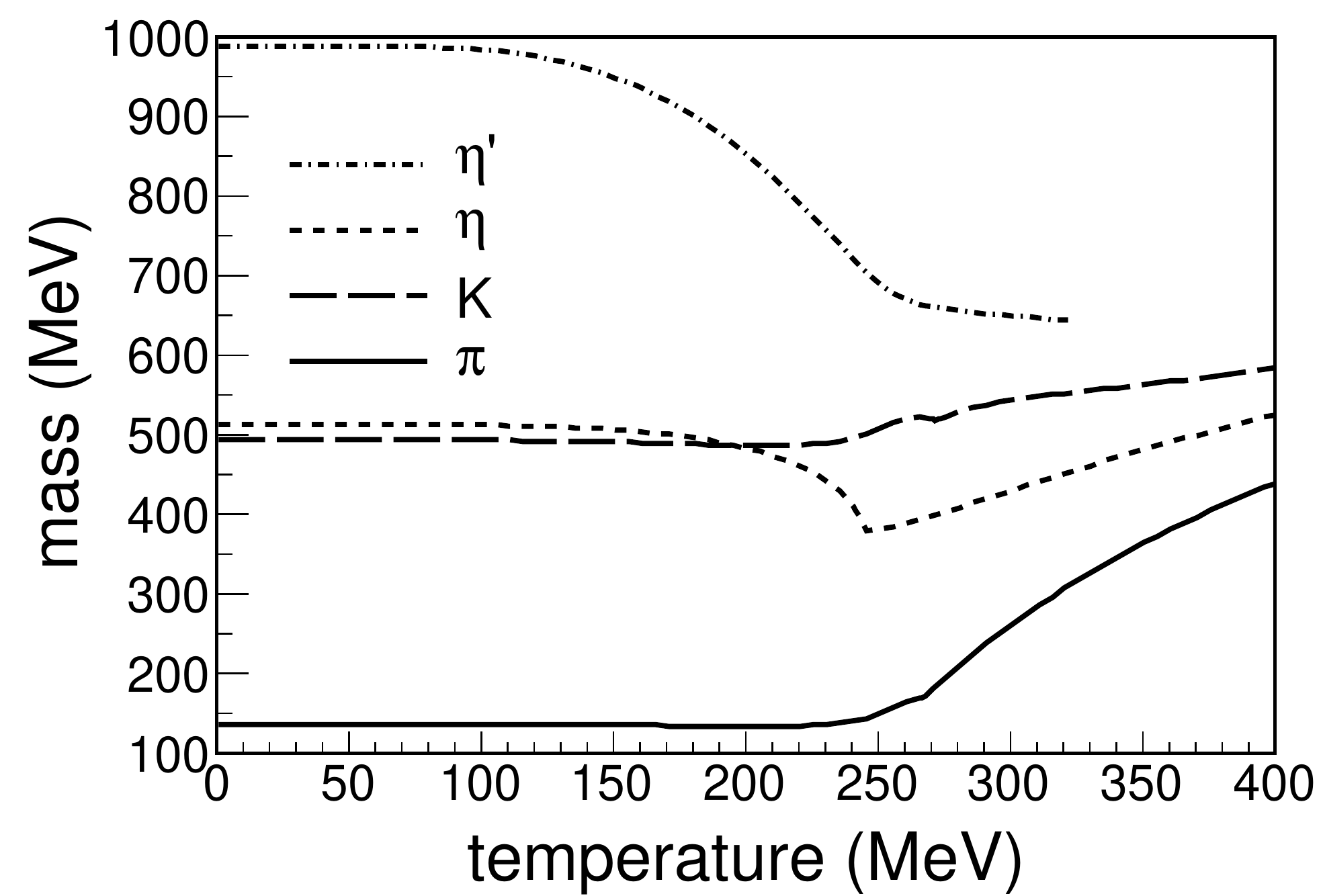}
\includegraphics[scale=0.4]{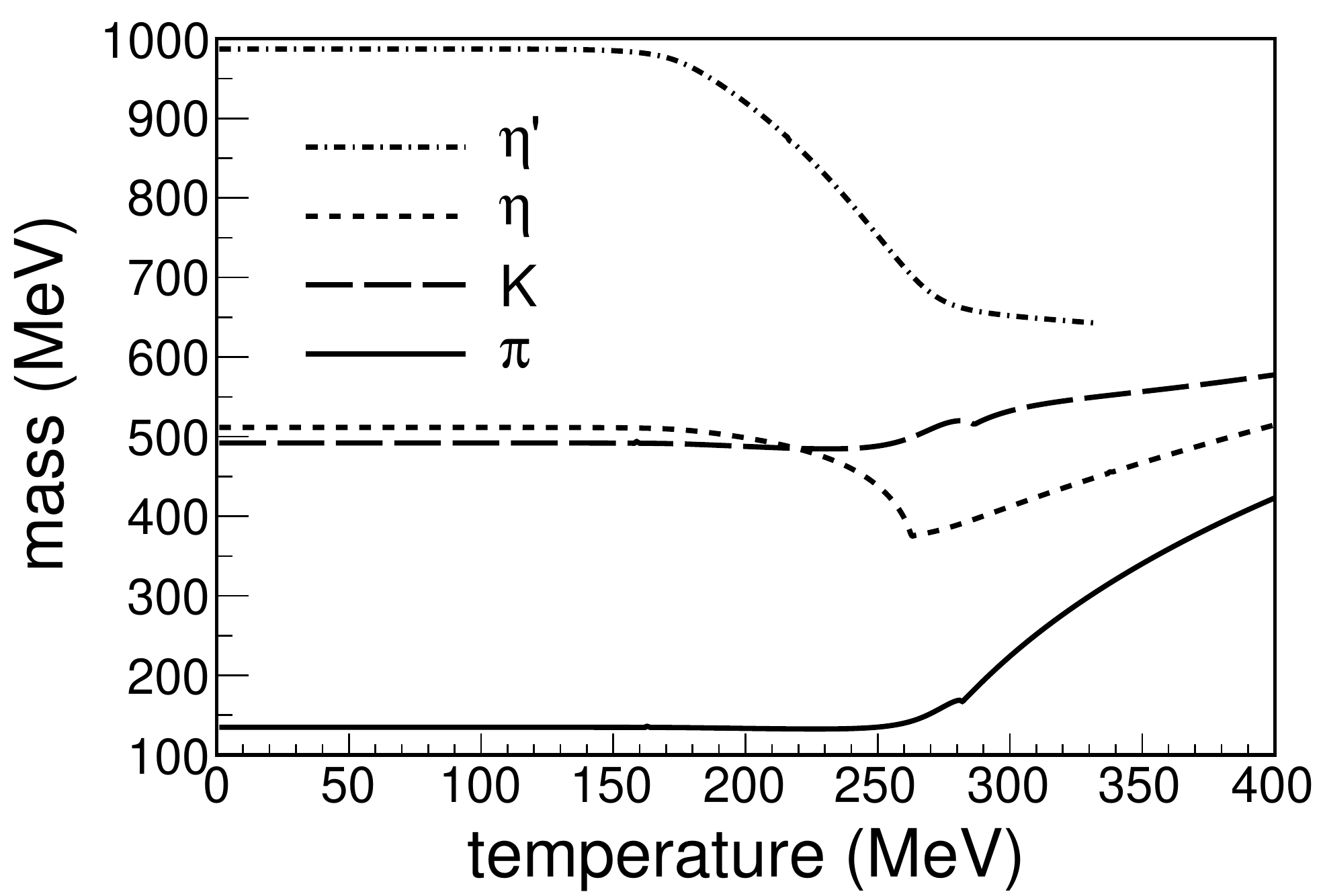}
\caption{\label{fig:Pmesons} Pseudoscalar meson masses as a function of the temperature 
for vanishing chemical potential in the NJL (left panel) and PNJL (right panel) models.}
\end{center} 
\end{figure*}

\begin{figure*}[htp]
\begin{center}
\includegraphics[scale=0.4]{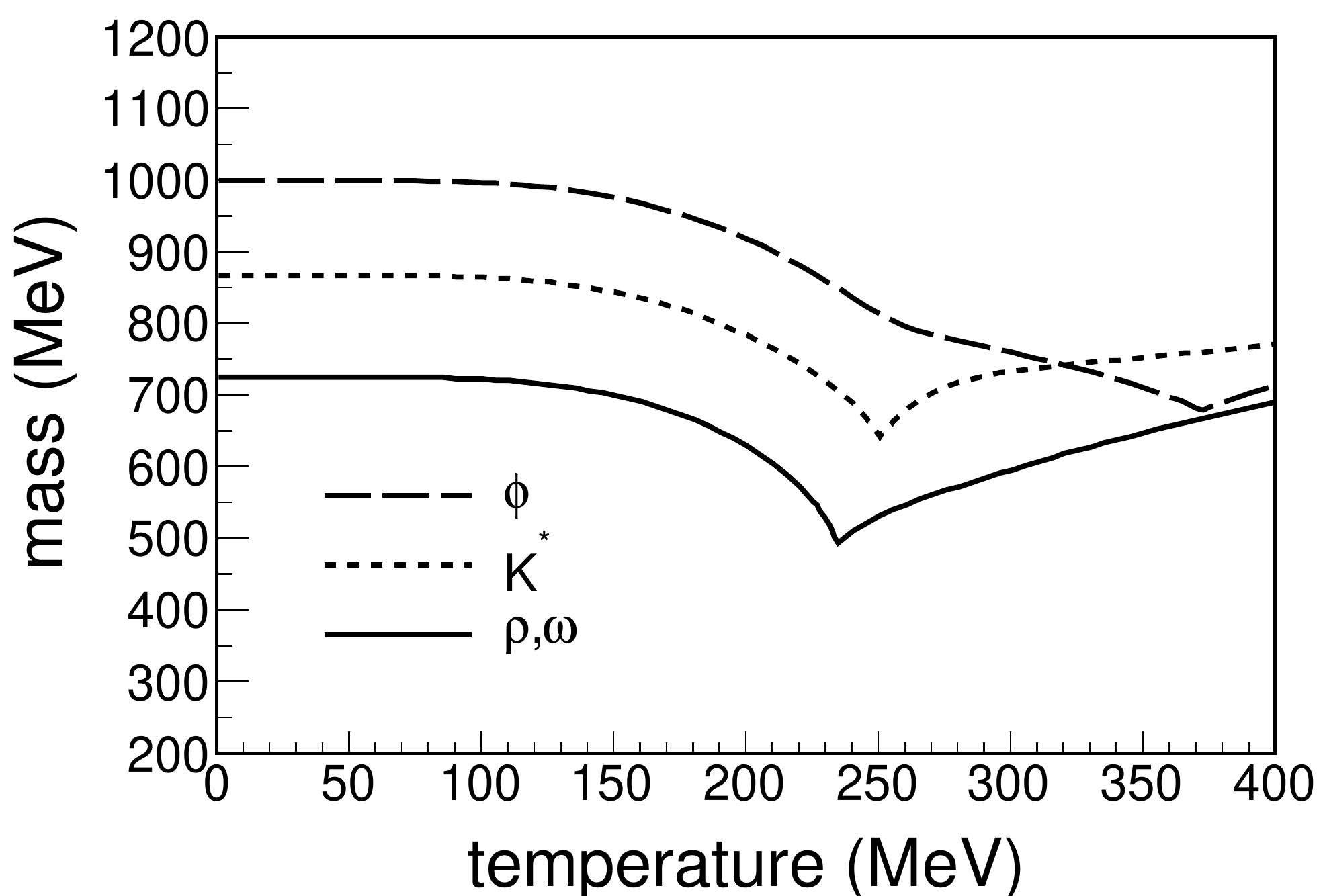}
\includegraphics[scale=0.4]{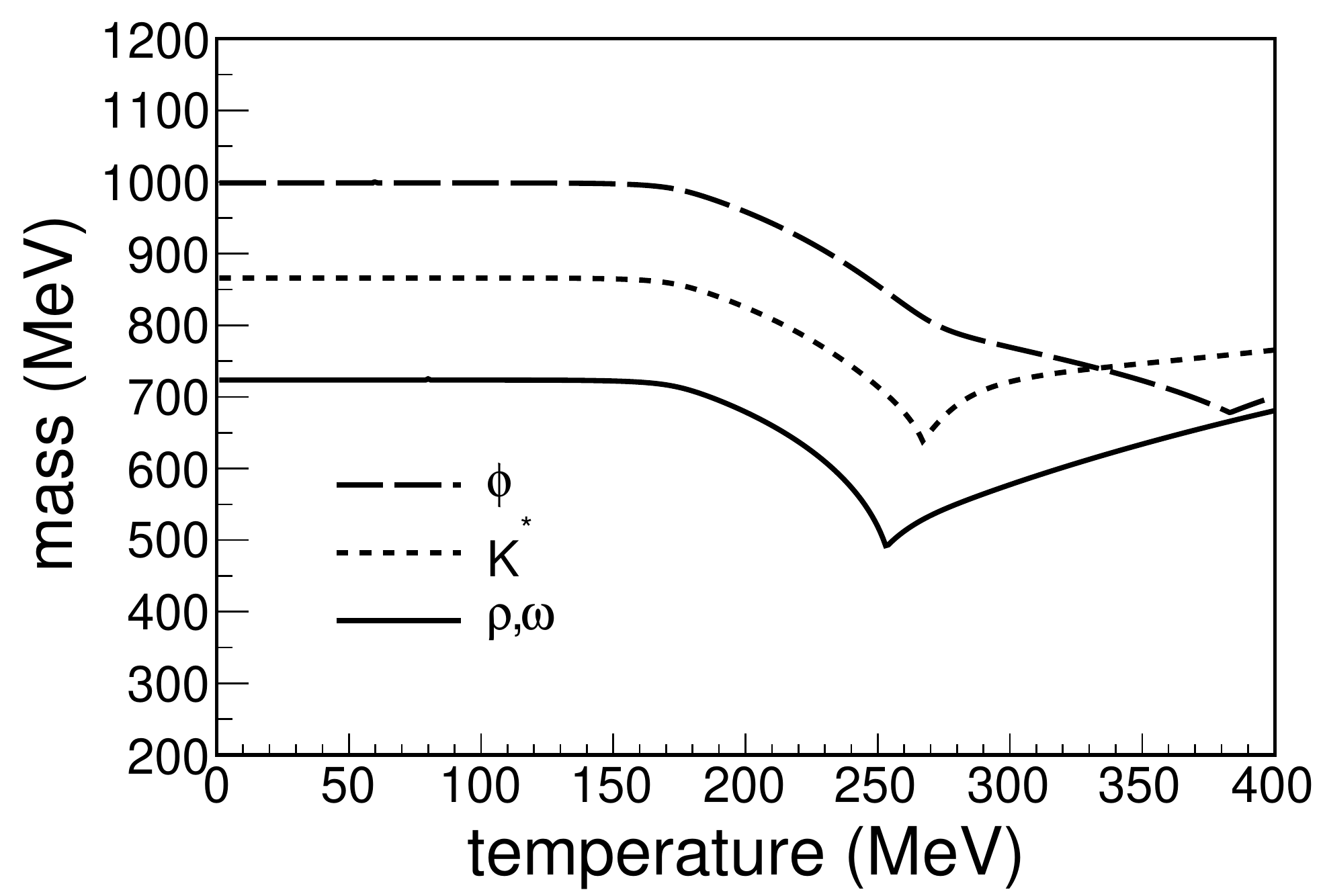}
\caption{\label{fig:Vmesons} Vector meson masses as a function of the temperature 
for vanishing chemical potential in the NJL (left panel) and PNJL (right panel) models.}
\end{center} 
\end{figure*}

  Our results for pseudoscalar and vector meson masses are summarized in Figs.~\ref{fig:Pmesons} and~\ref{fig:Vmesons}
respectively, where we include the results from both the NJL and the PNJL models.

In Table~\ref{tab:mesons} we present the masses of the pseudoscalar and vector mesons at zero temperature, as
well as the Mott temperatures for all of them, defined as the temperature at which
\be m_M (T_{Mott}) = m_p (T_{Mott}) + m_{{\bar q}} (T_{Mott}) \ , \ee
with $m_p$ and $m_{{\bar q}}$ the masses of the quark and antiquark that compose the meson.
\begin{table}
\begin{center}
\begin{tabular}{|c|c|c|c|c|c|c|c|c|}
\hline
Meson & $\pi$ & $K$ & $\eta$ & $\eta'$ & $\rho$ & $K^*$ & $\omega$ & $\phi$  \\
\hline
Mass at $T=0$ & 135 & 492 & 512 & 987 & 723 & 866 & 723 & 998  \\
Mass - pdg~\cite{Beringer:1900zz} & 136 & 495 & 548  & 958 & 775 & 892 & 782  & 1020  \\
\hline
$T_{Mott}$ - NJL & 267 & 271 & 245 & 0 & 234 & 250 & 234 & 372  \\
$T_{Mott}$ - PNJL & 282 & 286 & 262 & 0 & 253 & 266 & 253 & 382  \\
\hline
\end{tabular}
\caption{\label{tab:mesons} Masses at $T=0$ and Mott temperatures for the pseudoscalar and vector mesons in the 
NJL and PNJL models. For the $\eta'$ meson we find a finite decay width already at $T=0$. All units are given in MeV.}
\end{center}
\end{table}

  Notice that for each mesonic state a different Mott temperature is found, note also the large difference between the Mott temperatures within
the NJL and the PNJL models. Given the uncertainties of our model one could argue that a common Mott temperature may
work for them all. Alternatively one might claim that more precise data are necessary to establish an eventual difference of the Mott temperature of pions and kaons, for instance.
However, note that the $\phi$ meson has a very large Mott temperature, not consistent with the rest.
This fact makes the $\phi$ meson impossible to accommodate in a global picture of common hadronization conditions. This would manifest itself in a
larger $\phi/\pi$ ratio than the expected for a production at a common temperature, as seen experimentally in Ref.~\cite{Abelev:2014uua}.

\section{\label{sec:diquarks}Diquarks}

  A second Fierz transformation can convert the original NJL interaction into Lagrangian that describes the interaction 
among quarks~\cite{Buballa:2003qv}. Their bound states, diquarks, belong to a nonsinglet color representation and are
not experimentally observable states. However, they will be important for the construction of baryons.

   It is important to fix the different diquark channels we consider. In color space, we neglect the sextet representation (${\bf 3}_c \otimes {\bf 3}_c = \bar{
{\bf 3}}_c \oplus {\bf 6}_c$) because the members of this representation cannot be combined with an additional quark to form
colorless baryons (in addition, this channel is known to be repulsive).
In flavor space, diquarks from both ${\bf {\bar 3}}_f$ and ${\bf 6}_f$ representations can take part in the baryon structure, but they should be appropriately combined with the 
spin structure to have a total antisymmetric wavefunction~\cite{Vogl:1991qt, Reinhardt:1992}. A summary of the different allowed channels
is shown in Table~\ref{tab:diquarks}. All the allowed combinations can be alternatively
obtained by applying a Fierz transformation to the original color-current Lagrangian into the $qq$ sector~\cite{Buballa:2003qv}.
All the terms emerging from the Fierz transformation, exactly match all the different terms shown in
Table~\ref{tab:diquarks} (see further discussion and final Lagrangian in App.~\ref{app:fierz}).

\begin{table}
\begin{center}
\begin{tabular}{|c|c|c|c|c|}
\hline
Color & Flavor & $J^{P}$ & $\Gamma$  & Denomination \\
\hline
\hline
${\bf 6}_S$ & \multicolumn{3}{|c|}{Not considered here} & \\ \cline{2-5} 
${\bf \bar{3}}_A$  & ${\bf 6}_S$ & $1^+ $ & $\gamma_\mu$& Axial \\
${\bf \bar{3}}_A$  & ${\bf \bar{3}}_A$ & $0^+$ &$i \gamma_5$& Scalar \\
${\bf \bar{3}}_A$  & ${\bf \bar{3}}_A$ & $0^-$ &$\unit$& Pseudoscalar \\
${\bf \bar{3}}_A$  & ${\bf \bar{3}}_A$ & $1^-$ &$\gamma_\mu \gamma_5$ & Vector \\
\hline
\hline
\end{tabular}
\caption{\label{tab:diquarks} Different diquarks belonging to different sectors. $\Gamma$ denotes the spin structure associated 
with the $qq$ vertex.}
\end{center}
\end{table}

   As we will see later, only the low-lying diquarks of each spin (scalar and axial-vector ones) will be
used to form baryons. This is so due to the fact that the masses of the pseudoscalar and vector diquarks will be
higher than the experimental baryon masses and already unstable at zero temperature.
For this reason we only detail here the Lagrangian describing the scalar diquark sector
\be \label{eq:lagdiq} {\cal L}_{qq} = G_{DIQ} \ (\bar{\psi} i\gamma_5 C \tau^A T^{A'} \bar{\psi}^T) 
(\psi^T C i \gamma_5 \tau^{A} T^{A'} \psi) \ , \ee
and the one for the axial diquark sector
\be \label{eq:lagdiq1} {\cal L}_{qq} = G_{DIQ,V} \ (\bar{\psi} \gamma^\mu C \tau^S T^{A'} \bar{\psi}^T) 
(\psi^T C  \gamma_\mu \tau^{S} T^{A'} \psi) \ , 
\ee
where $G_{DIQ}$ and $G_{DIQ,V}$ are coupling constants (related to the original $g$ but taken here as a free parameters) and
$C=i\gamma_0 \gamma_2$ represents the charge-conjugation operation. Finally, we have denoted by $\tau^A$ and $\tau^S$
the antisymmetric and symmetric flavor matrices, respectively; and by $T^{A'}$ the antisymmetric color matrices. In
particular, the presence of the latter reflects that the diquarks cannot be color singlets.

   The BS equation for the quark-quark scattering in the RPA approximation reads 
\begin{widetext}
\be \label{eq:BSdiquark} T^{ab}_{ij,mn} (p^2) = {\cal K}^{ab}_{ij,mn} + i \int \frac{d^4 k}{(2 \pi)^4}
{\cal K}^{ac}_{ij,pq} \  S_{p} \left( k+ \frac{p}{2} \right) \ S^c_{q}  \left( \frac{p}{2}-k \right) \ T^{cb}_{pq,mn} (p^2) \ , \ee
\end{widetext}
where $S^c(p) \equiv C^{-1} S^T(-p) C$ denotes the charge-conjugated quark propagator, with $T$ the transposed operator (not to be confused with
the temperature).

  Details concerning the simplification of this equation are given in App.~\ref{app:bs}. In terms of the function
$t^{ab}(p^2)$, 
\be T_{i j,m n}^{ab} (p^2) = \Omega^a_{ij} \ t^{ab} (p^2) \ \bar{\Omega}^b_{nm} \ , \ee
we can express the solution of the BS equation as
\be t^{ab} (p^2) = \frac{2 G_{DIQ} }{1- 2  G_{DIQ} \Pi^{ab} (p^2) } \ , \ee
with the quark-quark polarization function of Eq.~(\ref{eq:poldiq}).

   For the spin-1 channels the polarization function contains a transverse and a longitudinal terms
\be \label{eq:components} \Pi_{\mu \nu}^{ab} = \Pi^{ab}_\perp {\cal T}_{\mu \nu} + \Pi^{ab}_\parallel {\cal L}_{\mu \nu} \ , \ee
where we have defined the projectors,
\be {\cal T}_{\mu \nu} = g_{\mu \nu}- \frac{p_\mu p_\nu}{p^2} \ , \quad {\cal L}_{\mu \nu} = \frac{p_\mu p_\nu}{p^2} \ . \ee
The solution of the spin-1 diquark masses involves the transverse component of the polarization function
$\Pi^{ab}_\perp (p^2) = \frac{1}{3} {\cal T}^{\mu \nu} \Pi_{\mu \nu}^{ab}$.
For axial diquarks this fact directly comes from the vector current conservation, but for vector diquarks this result
still holds~\cite{Klevansky:1997dk}.
 
The function $t^{ab} (p^2)$ reads
\be t^{ab}(p^2) = \left[ \frac{2 G_{DIQ,V} }{1- 2  G_{DIQ,V} \ \Pi_\perp (p^2) } \right]^{ab} \ , \ee
where the coupling constant $G_{DIQ,V}$ for vector and axial diquarks is, in principle, related to $G_{DIQ}$ by the Fierz transformation.
However, we will take it here as a free parameter to be fixed by a fit to the baryon masses.

   Again, the poles of the $t^{ab} (p^2)$ functions are identified with dynamically generated diquarks in their
respective flavor and spin channel. The diquark mass, $m_{DIQ}$, is obtained by the solution of the equation 
$1- 2  G_{DIQ} \Pi^{ab} (m_{DIQ}^2)=0$ for spin zero diquarks and
 $1- 2 G_{DIQ,V} \ \Pi^{ab}_\perp (m_{DIQ}^2)=0$ for spin one diquarks.

\begin{figure}[htp]
\begin{center}
\includegraphics[scale=0.42]{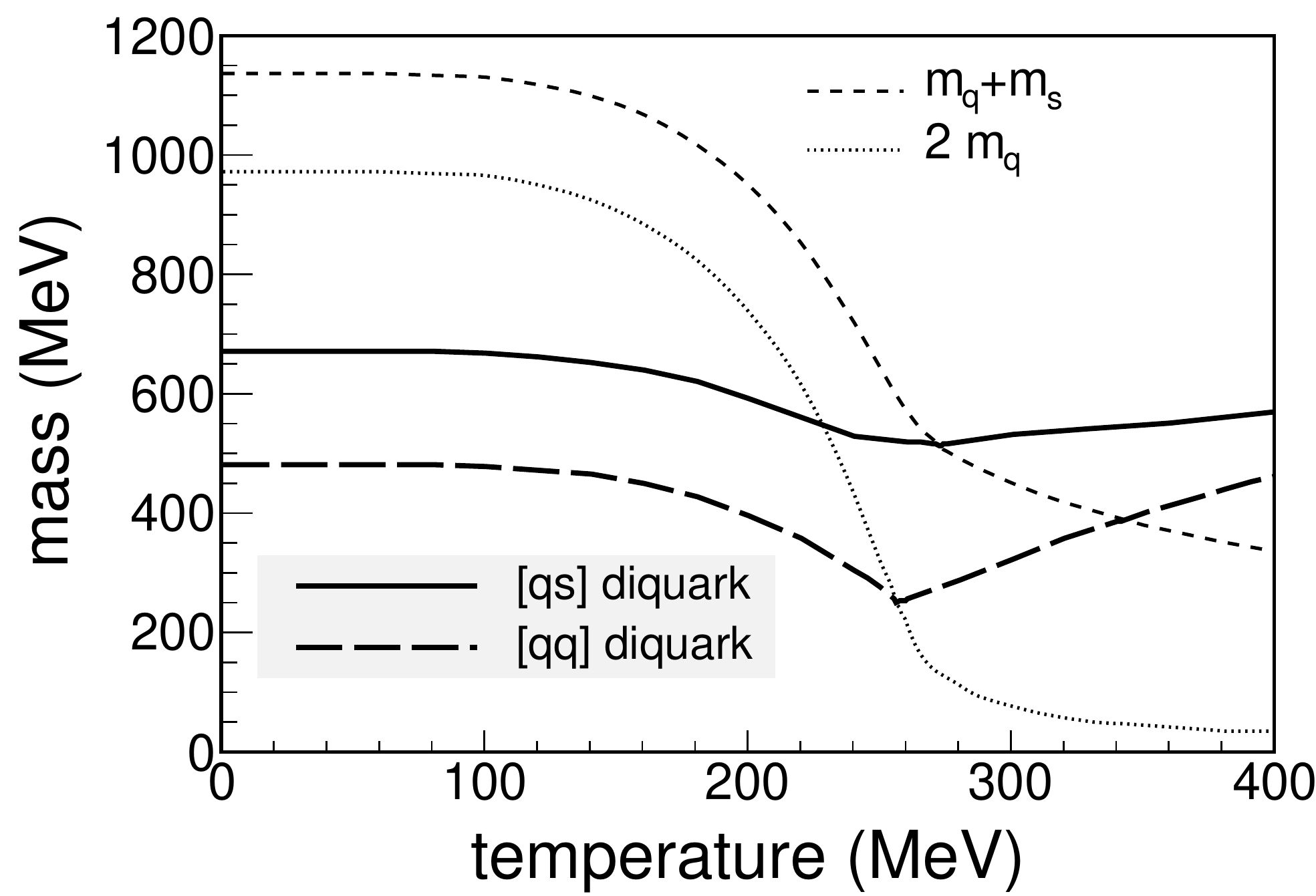}
\includegraphics[scale=0.42]{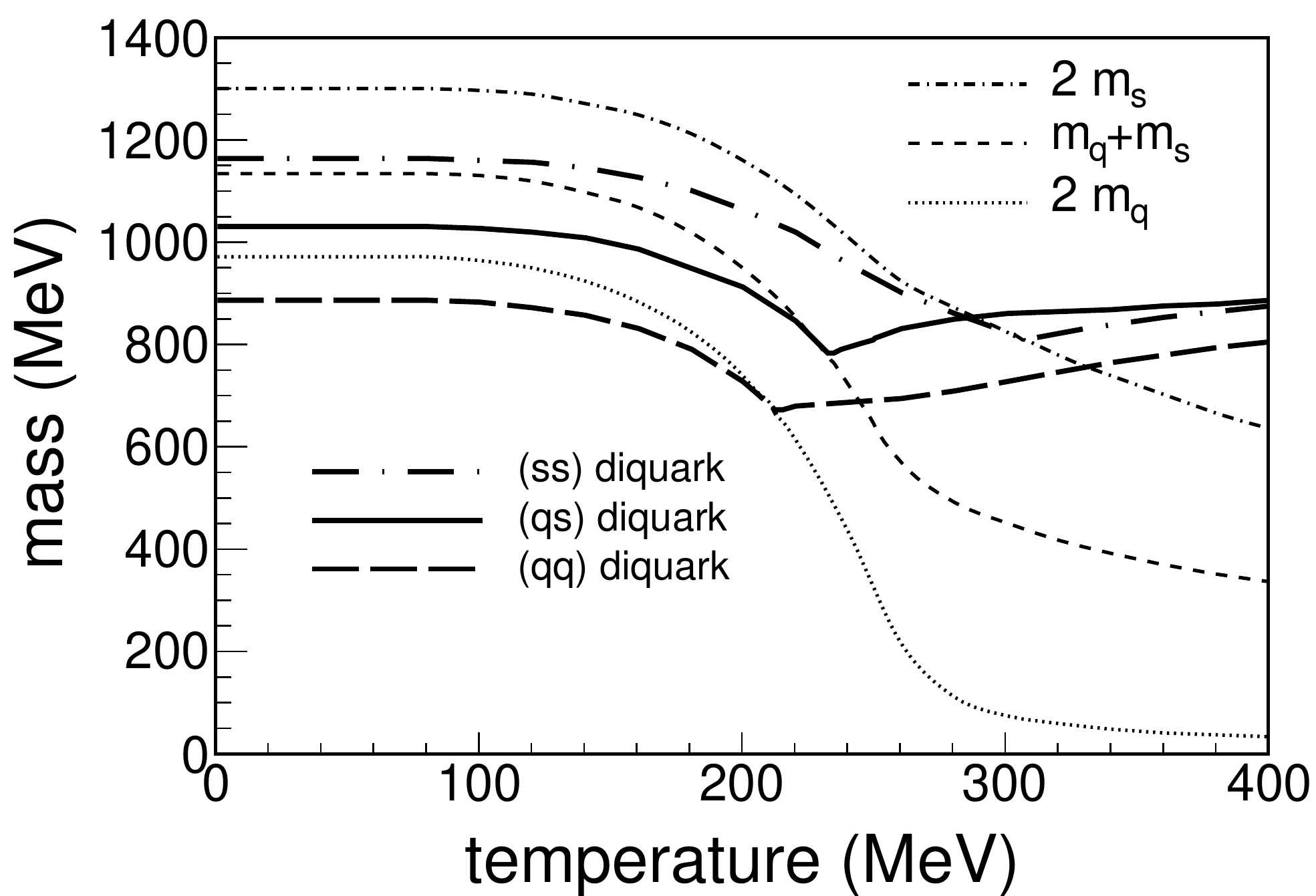}
\caption{\label{fig:diq_sv} Scalar diquark (upper panel) and axial diquark (lower panel) masses as a function of
the temperature for vanishing chemical potential in the NJL model. For comparison, the thermal quark masses are
also shown. $q$ stands for light quark ($q=u=d$).}
\end{center} 
\end{figure}
   
   In Fig.~\ref{fig:diq_sv} we present our results for scalar and axial diquark masses, which will be used 
to model baryons in the next section. Scalar diquarks are represented by square brackets $[q_1q_2]$ and axial diquarks 
by parenthesis $(q_1q_2)$. From Fig.~\ref{fig:diq_sv} it is possible to read off the Mott temperature for
the different states (defined as the temperature at which the mass of the bound state equals the sum of
the quark masses). Beyond this temperature, a diquark thermal width is generated, which represents the probability
of the diquark to decay into a pair of quarks. 

\begin{figure}[htp]
\begin{center}
\includegraphics[scale=0.42]{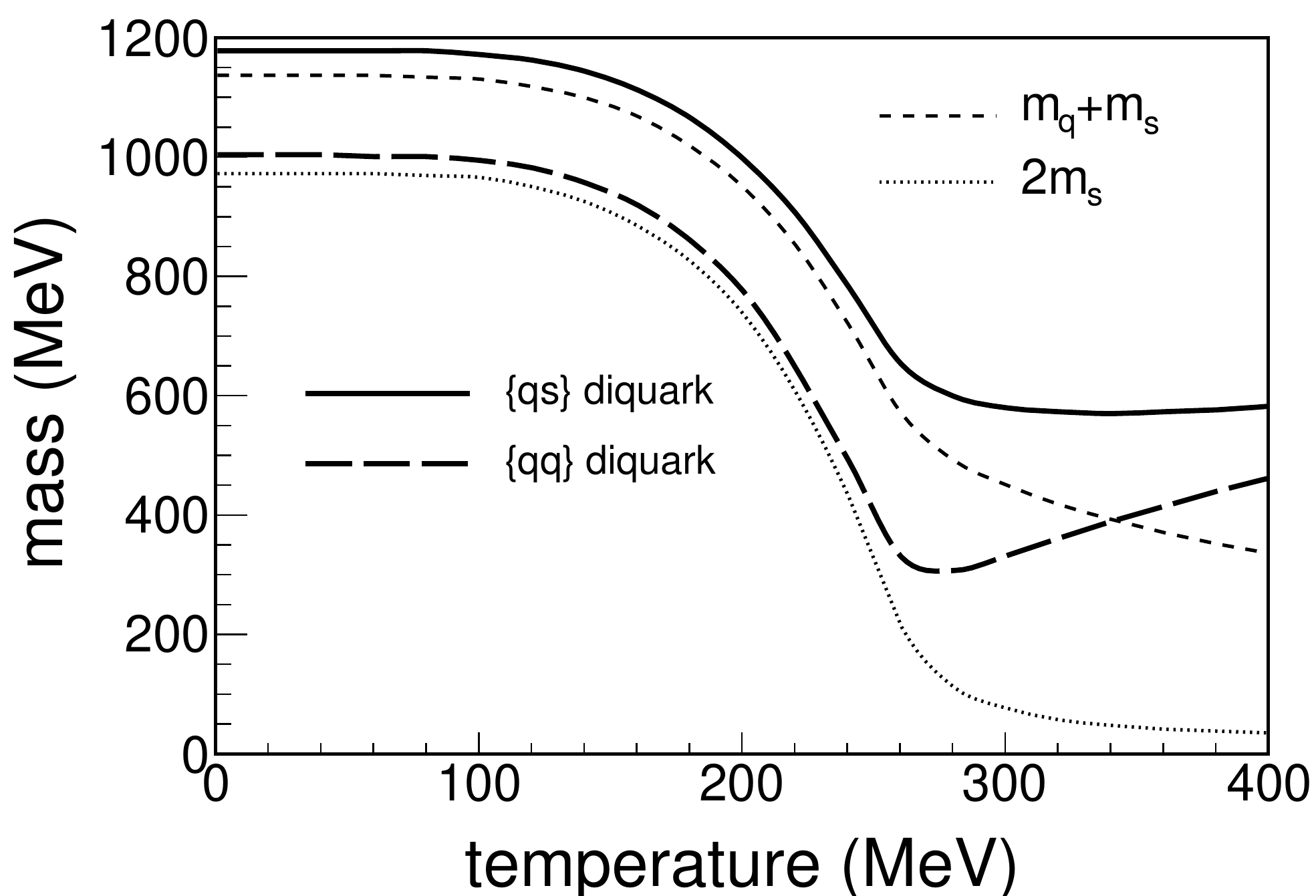}
\includegraphics[scale=0.42]{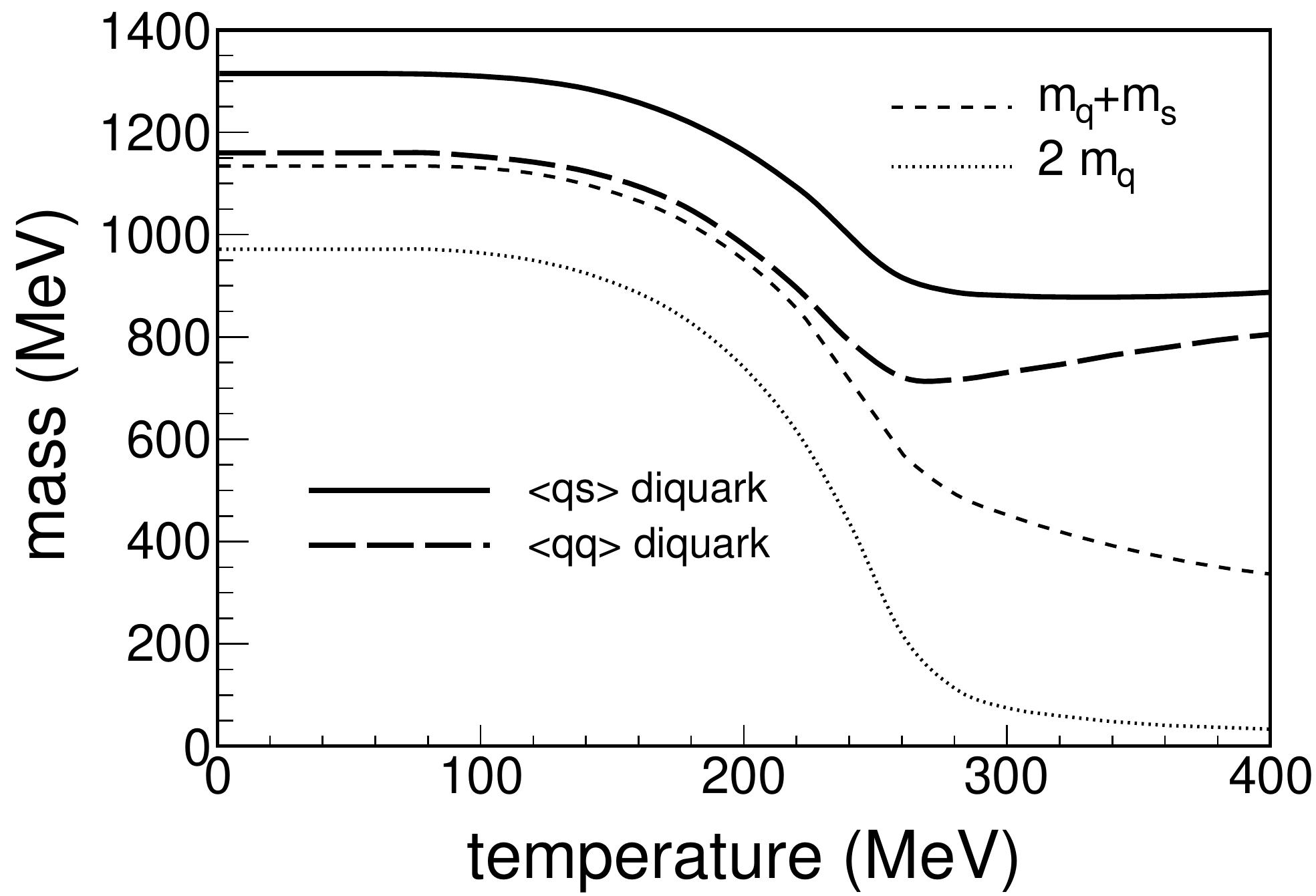}
\caption{\label{fig:diq_pa} Pseudoscalar diquark (upper panel) and vector diquark (lower panel) masses as
a function of temperature for vanishing chemical potential in the NJL model. Notice that the diquark masses are
always higher than the combined mass of their constituents. Therefore, they are unstable against decay to two quarks.}
\end{center}
\end{figure}

For completeness, we also present our results for the pseudoscalar $\{q_1q_2\}$ and vector $<q_1q_2>$ diquarks in
Fig.~\ref{fig:diq_pa}. As we have anticipated, at $T=0$ these states have a finite thermal decay width which
excludes a role in forming  stable baryons at low temperatures. For this reason, we will
neglect these states hereafter.

\begin{figure*}[htp]
\begin{center}
\includegraphics[scale=0.42]{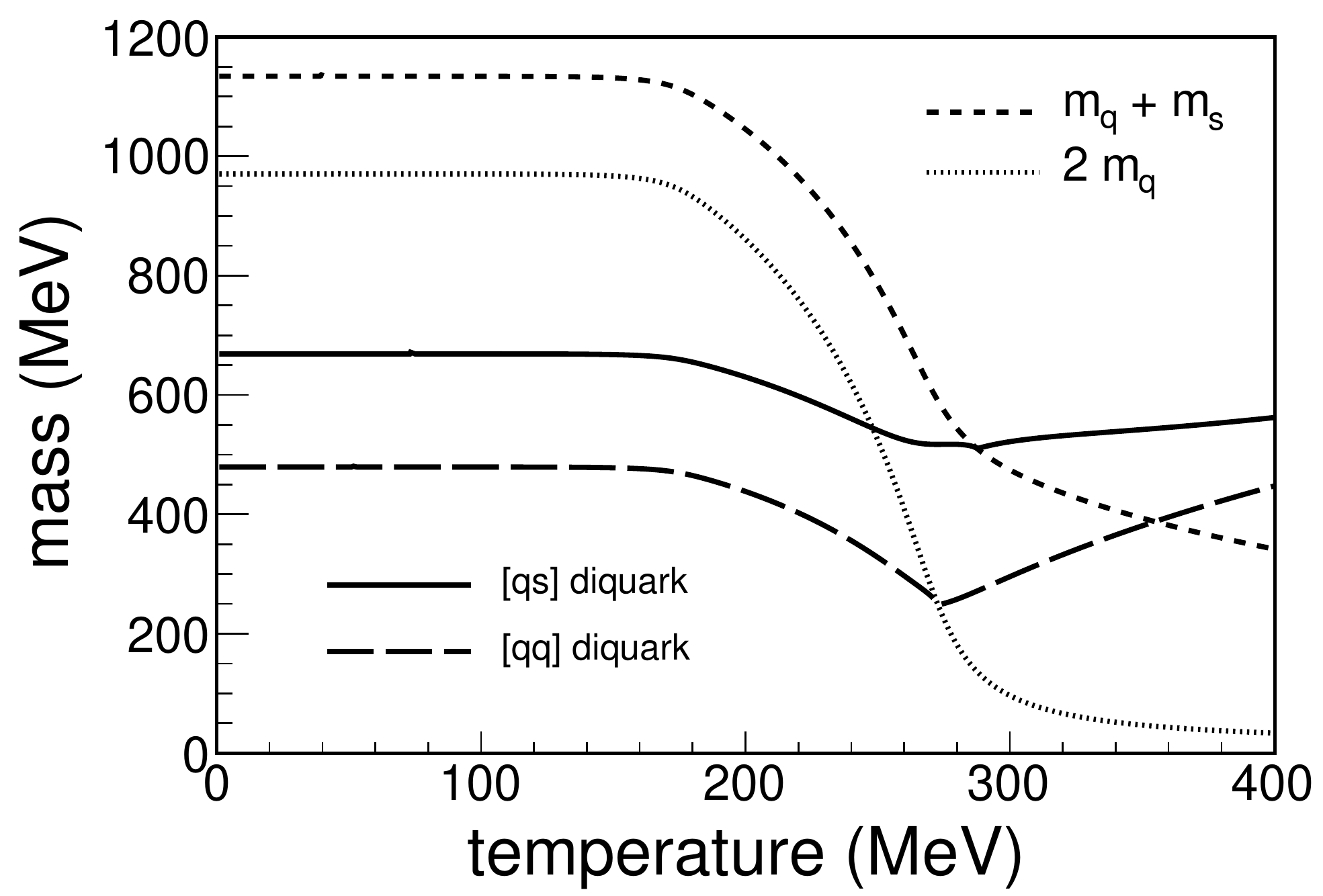}
\includegraphics[scale=0.42]{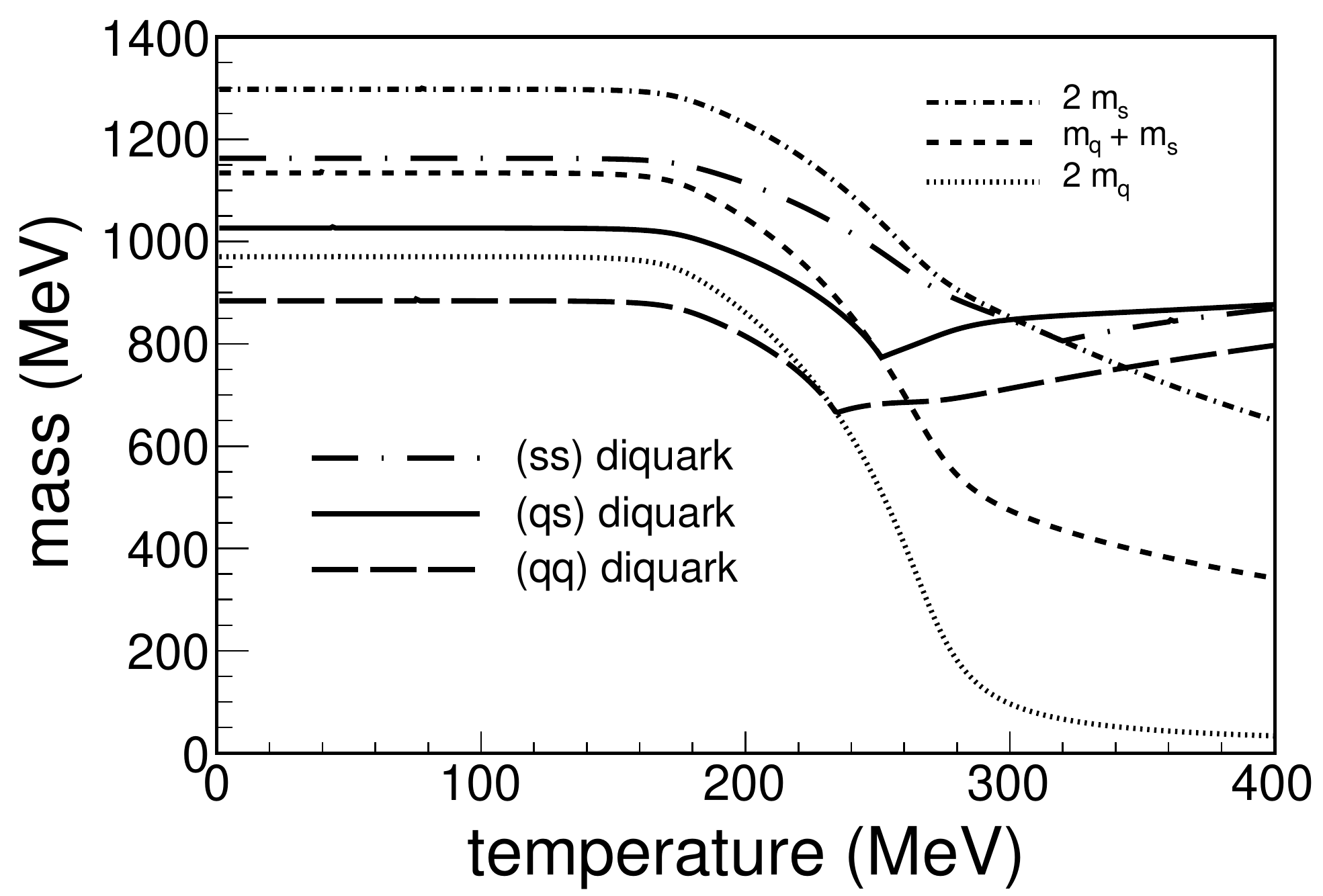}
\caption{\label{fig:diq_pnjl} Scalar diquark (left panel) and axial diquark (right panel) masses as a function
of temperature for vanishing chemical potential in the PNJL model.}
\end{center} 
\end{figure*}

  We now turn to the PNJL model and show the results for the scalar and axial diquark masses with the
parameter set presented in Table~\ref{tab:param}. In Fig.~\ref{fig:diq_pnjl} we show the diquark masses as a function of temperature at zero chemical
potential. In this case, the quark masses are more stable as a function of the temperature, generating a
systematically larger Mott temperature in comparison with the NJL model. At $T=0$ the masses obtained from
the NJL and PNJL models coincide, providing a consistency check.

   In Table~\ref{tab:mott_di} we present a summary of Mott temperatures (with precision of 1 MeV) for the different states in
the NJL and in the PNJL model. The PNJL model shows higher melting temperatures in all cases. From this table one 
already sees a hint for the flavor dependence of the hadronization (Mott) temperature. This temperature increases with the strangeness content of the diquark.

\begin{table}
\begin{center}
\begin{tabular}{|c|c|c|}
\hline
Diquark & NJL $T_{Mott}$ (MeV) & PNJL $T_{Mott}$ (MeV)\\
\hline
\hline
$[qq]$ & 256 & 272\\
$[qs]$ & 273 & 287 \\
$(qq)$ & 212 & 234 \\
$(qs)$ & 233 & 251 \\
$(ss)$ & 307 & 319 \\
\hline
\end{tabular}
\caption{\label{tab:mott_di} Mott (or melting) temperature for the different scalar $[q_1q_2]$ and vector axial $(q_1q_2)$ diquarks in 
the NJL as well as in the  PNJL model.}
\end{center}
\end{table} 

   Assuming the pole dominance of the diquark propagation, we can expand the $t^{ab}(p^2)$ function around its pole to obtain~\cite{Buballa:2003qv}:
\be \label{eq:propscalar} t^{ab} (p^2) = -\frac{g^2_{[qq] \rightarrow qq}}{p^2-m_{DIQ}^2 } \ . \ee
Taking the diquark to be at rest (${\bf p}=0$) the effective coupling $g^2_{[q_1q_2] \rightarrow q_1q_2}$ is defined as
\be\label{eq:geffS} g^2_{[q_1q_2] \rightarrow q_1q_2} = \frac{2m_{DIQ}}{\left. \frac{\pa \Pi^{ab}(p_0)}{\pa p_0} \right|_{p_0=m_{DIQ}} } \ . \ee
For the axial diquarks we find in the pole approximation
\be \label{eq:propaxial} t^{ab,\mu \nu} (p_0,0) = \frac{g^2_{(q_1q_2) \rightarrow q_1q_2}}{p_0^2-m_{DIQ}^2 } \left( g^{\mu \nu} - \frac{p^\mu p^\nu}{m_{DIQ}^2}\right) \ , \ee
with
\be \label{eq:geffA} g^2_{(q_1q_2)\rightarrow q_1q_2} = \frac{-2 m_{DIQ}}{\left. \frac{\pa \Pi^{ab}_{\perp}(p_0)}{\pa p_0} \right|_{p_0=m_{DIQ}} } \ . \ee

\begin{figure*}[htp]
\begin{center}
\includegraphics[scale=0.42]{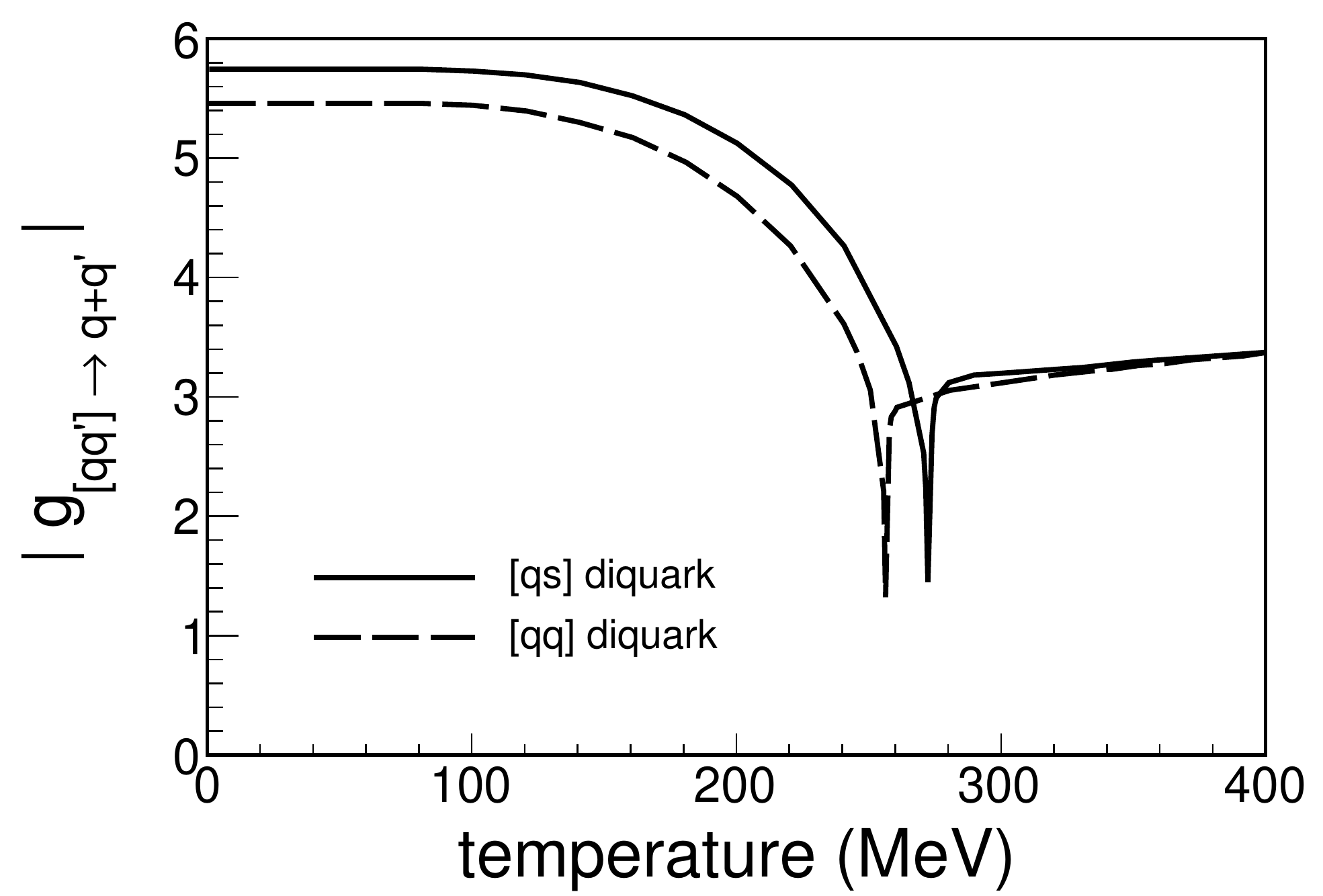}
\includegraphics[scale=0.42]{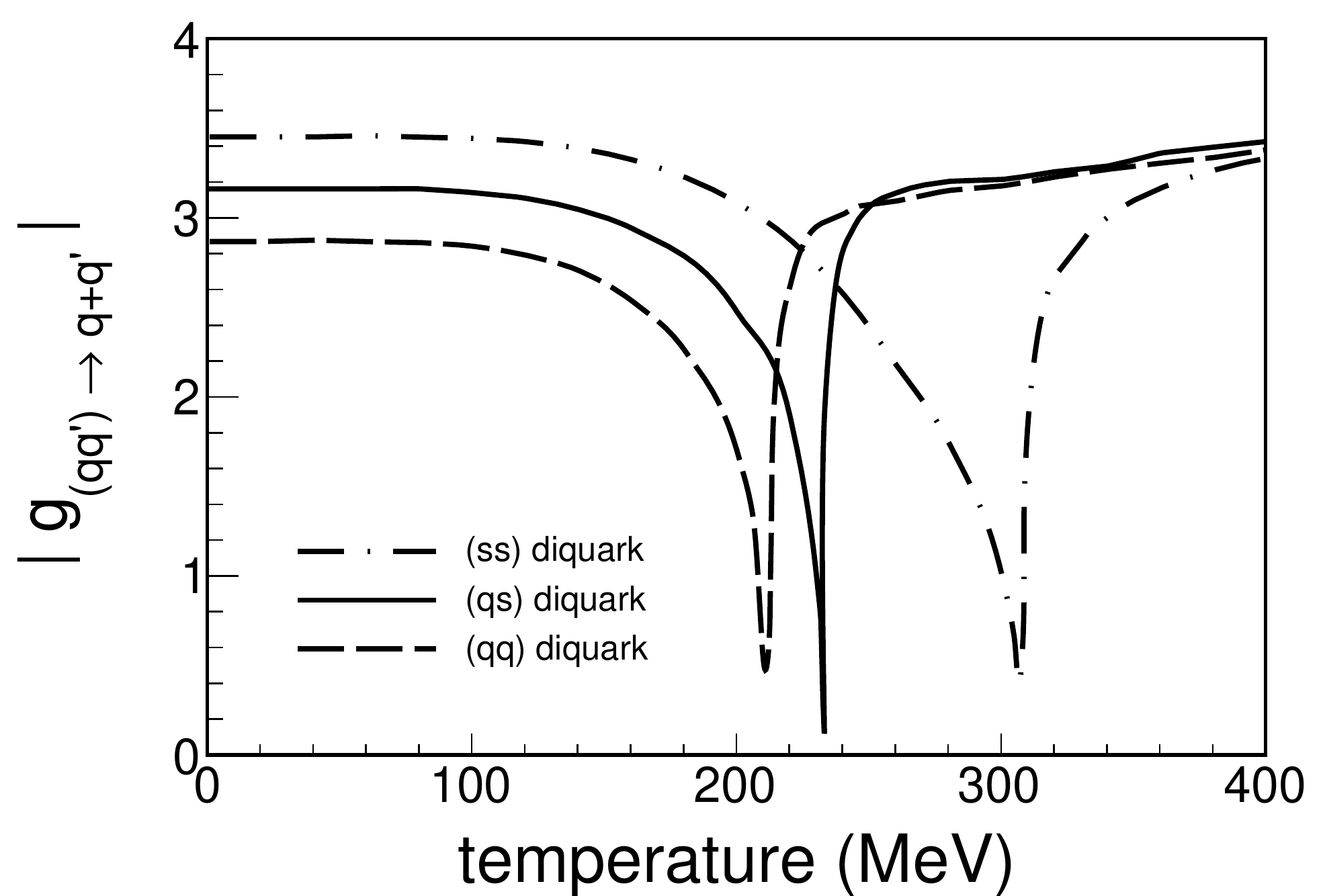}
\caption{\label{fig:geff_NJL} Effective diquark-quark-quark coupling, defined in Eq.~(\ref{eq:geffS}) for the scalar diquarks (left panel)
and in Eq.~(\ref{eq:geffA}) for the axial diquark (right panel) as a function of temperature in the NJL model.}
\end{center} 
\end{figure*}

\begin{figure*}[htp]
\begin{center}
\includegraphics[scale=0.42]{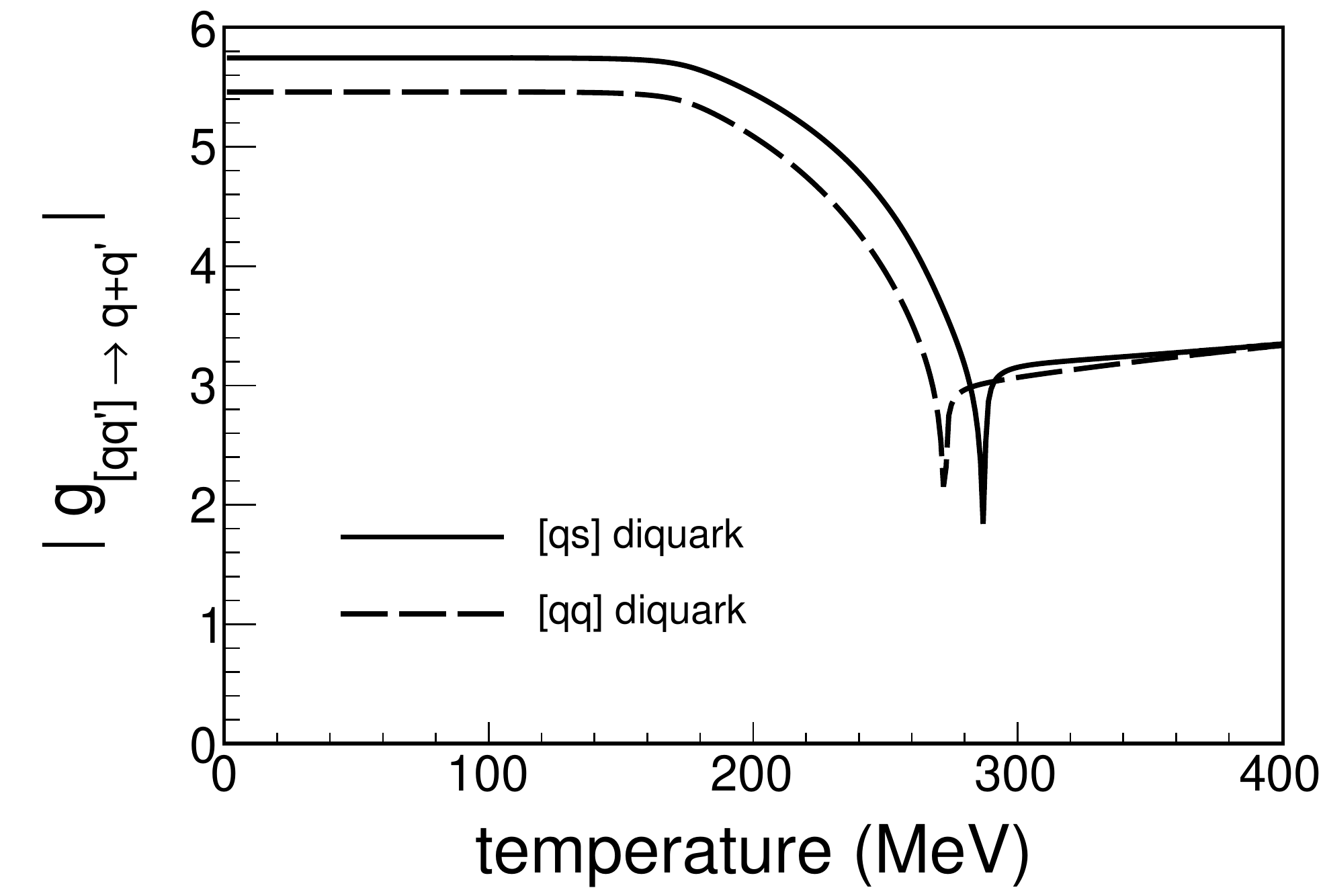}
\includegraphics[scale=0.42]{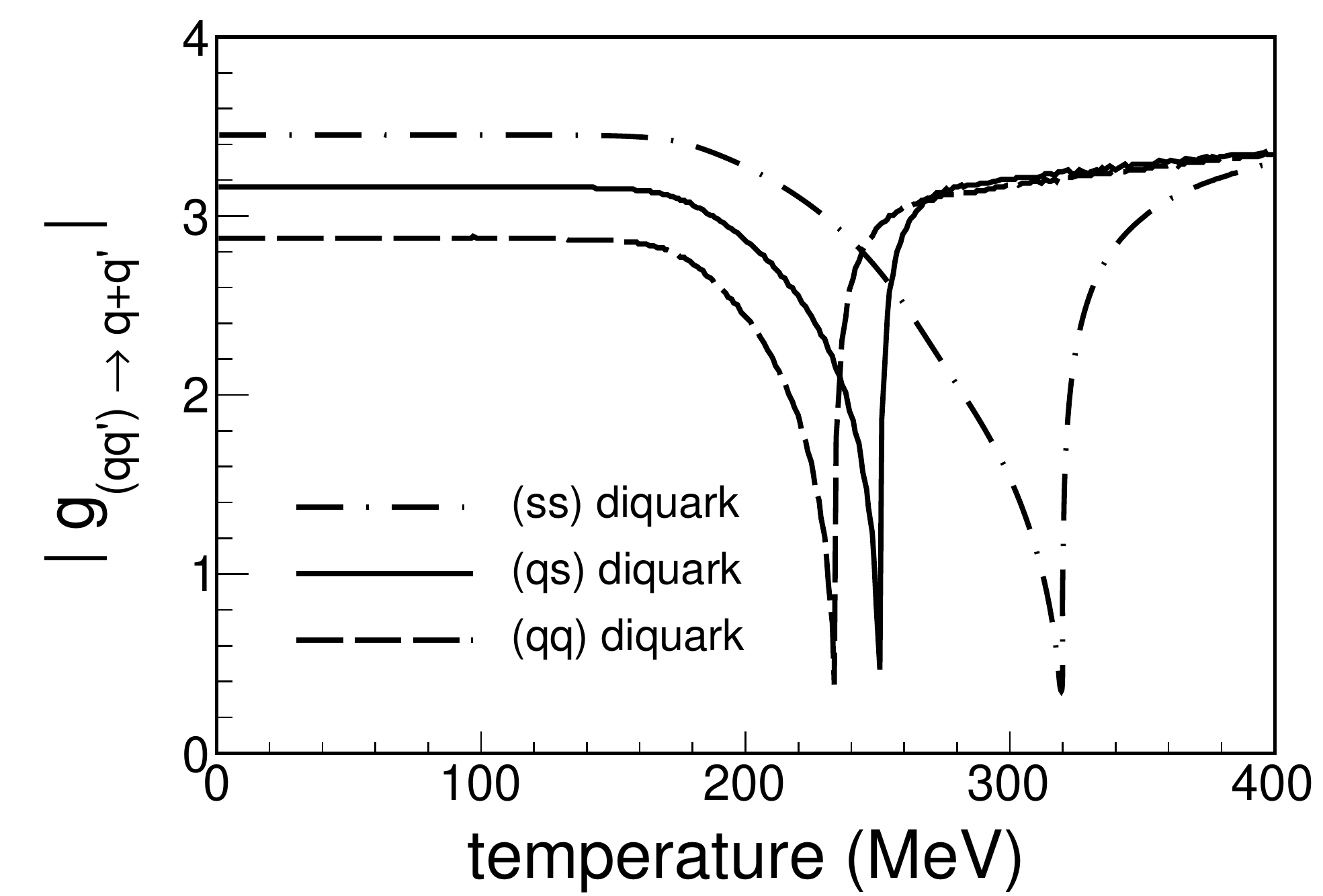}
\caption{\label{fig:geff_PNJL} Same as Fig.~\ref{fig:geff_NJL} but for the PNJL model.}
\end{center} 
\end{figure*}

   We now present our results for the effective couplings of the scalar and axial diquarks as a function of the temperature
at vanishing chemical potential. The NJL results are given in Fig.~\ref{fig:geff_NJL} and those for the PNJL model in Fig.~\ref{fig:geff_PNJL}.
Notice that the Mott temperature clearly coincides in these plots with the value at which the effective coupling vanishes.

\section{\label{sec:baryon} Quark-diquark bound states}

  In the last two sections we have explored the concept of ``hadronization'' as a dynamical generation of bound
states from quark and antiquark scattering. This idea --quite successful for the generation of mesons and diquarks-- can
be pushed forward to describe baryons as bound states of a quark and a diquark. For $N_f=3$ we will consider
both the octet and decuplet flavor representations of baryons. Scalar diquarks (those belonging to flavor $\bar{{\bf 3}}$ representation) and
axial diquarks (${\bf 6}$) will be used to build up the baryon octet and decuplet states, respectively, according
to the decomposition,
\be {\bf 3} \otimes ( \bar{ { \bf 3}} \oplus {\bf 6}) = ({\bf 1} \oplus {\bf 8}) \oplus ({\bf 8} \oplus {\bf 10}) \ . \ee

Notice that the members of the octet also contain nonzero contribution from the axial diquarks. However, previous
results at zero temperature have shown that the contribution is almost negligible~\cite{Lehmann}. For simplicity, we
will neglect the axial diquark contribution to the baryon octet.

  The starting point to describe baryons is the Fadeev equation~\cite{Buck:1992wz,Ishii:1995bu}:
\begin{widetext}
\be \label{eq:fadeev}  \left.  X_j^{\bar{j} ,\alpha} (P^2,q) - \int \frac{d^4 k}{(2\pi)^4} 
L_{j k}^{\bar{j} \bar{k}, \alpha \beta}(P^2,q,k) X_{k}^{\bar{k} ,\beta} (P^2,k)\right|_{P^2=M^2_B}= 0 \ , \ee
\end{widetext}
where the baryon wave function is denoted by $X_j^{\bar{j}, \alpha}$ and it carries a quark index ($j$), diquark index ($\bar{j}$),
and a possible spin index $\alpha$.

\begin{figure*}[htp]
\begin{center}
\includegraphics[scale=0.7]{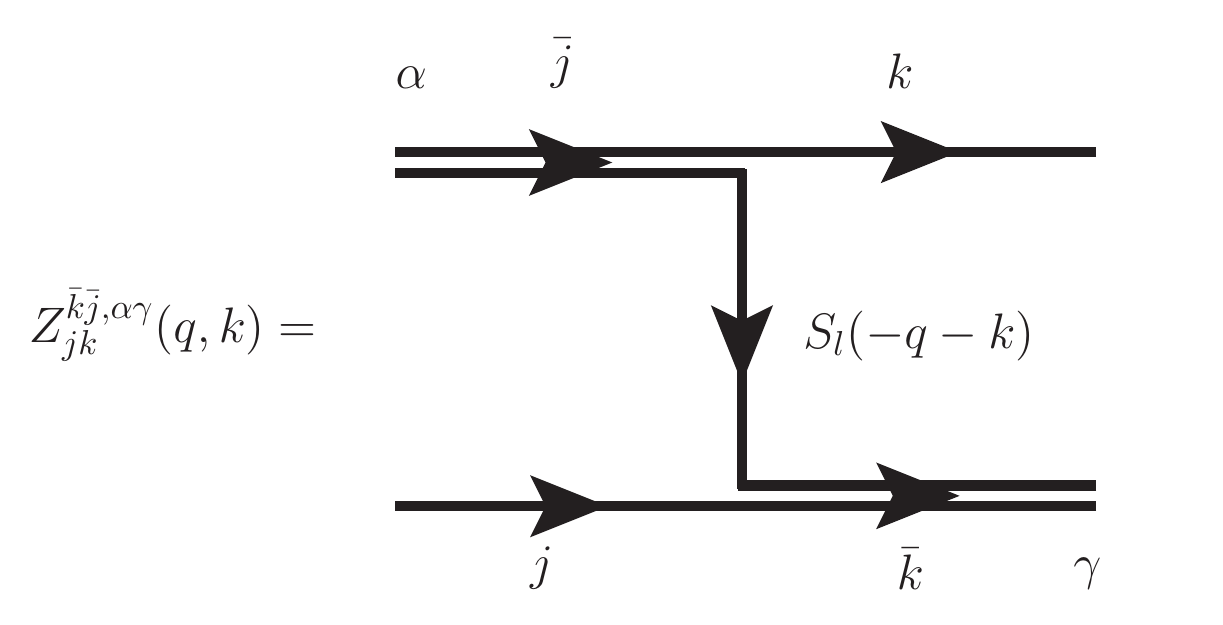}
\includegraphics[scale=0.75]{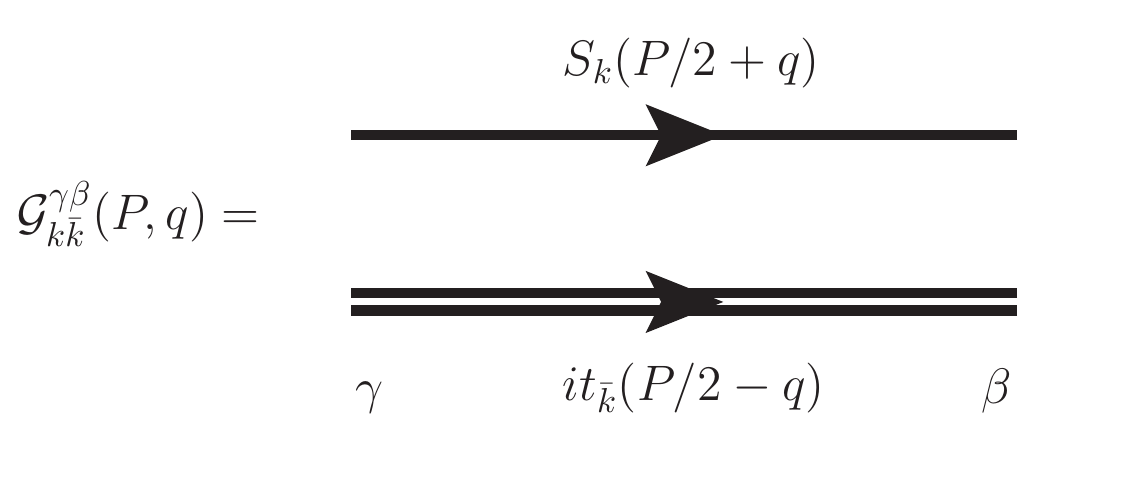}
\end{center} 
\caption{\label{fig:fadeev} Left panel: Effective coupling in the Fadeev equation which contains a quark exchange.
Right panel: Two-particle (quark+diquark) propagator in the Fadeev kernel.}
\end{figure*}

  The kernel reads~\cite{Buck:1992wz} 
\be \label{eq:kernel} L^{\bar{j}\bar{k} ,\alpha \beta}_{j k} (P^2,q,k) =  
\ {\cal G}_{k \bar{k}}^{\gamma \beta} (P^2,q)  Z_{jk}^{\bar{k} \bar{j},\alpha \gamma} (q,k) \ ,  \ee
with a first term which accounts for the free quark and diquark propagators (see right panel of Fig.~\ref{fig:fadeev})
\be {\cal G}_{k \bar{k}}^{\gamma \beta} (P^2,q) = S_k (P/2+q) \ it^{\gamma \beta}_{\bar{k}} (P/2-q) \ee
and a second term
\be Z_{jk}^{\bar{k} \bar{j},\alpha \gamma} (q,k)= \Omega_{jl}^{\bar{k},\gamma} \ S_l (-q-k) \ \Omega_{lk}^{\bar{j},\alpha} \ , \ee
which represents an interaction with an exchanged quark (displayed in the left panel of Fig.~\ref{fig:fadeev}).

  We do not attempt here to justify the form of the Fadeev equation (\ref{eq:fadeev}) and we refer the reader
to the original papers~\cite{Reinhardt:1992,Buck:1992wz} to learn the rigorous derivation and know more details.
 
\begin{figure*}[htp]
\begin{center}
\includegraphics[scale=0.5]{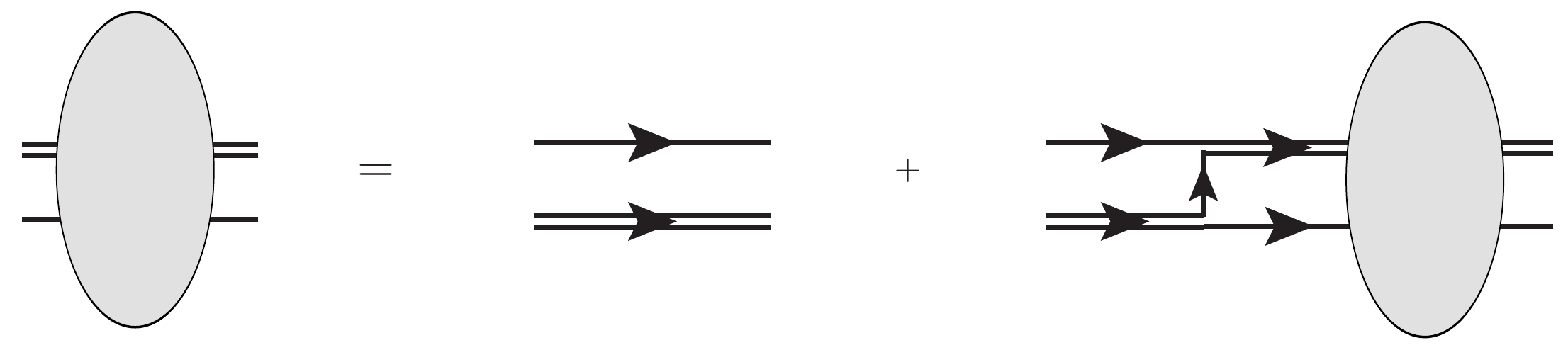}
\end{center} 
\caption{\label{fig:motiv} Dyson equation for the baryon propagator ${\bf G}= {\cal G}  + {\cal G}  Z {\bf G}$.}
\end{figure*}

  Nevertheless, we can provide a simple motivation for Eq.~(\ref{eq:fadeev}):
If we denote by ${\bf G} (P^2)$ the full baryon propagator, one can form a Dyson equation by taking
${\cal G}$ as the leading order approximation (free propagation), and then considering $Z$ as the elementary
interaction (see Fig.~\ref{fig:motiv}).
The Dyson equation can be symbolically written as
\be {\bf G}= {\cal G}  + {\cal G}  Z {\bf G} \ ,  \ee
whose solution reads
\be {\bf G} = \frac{ {\cal G}}{1- {\cal G} Z} \ . \ee
The baryon masses are now extracted as the poles of the baryon propagator, so one needs to solve ${\bf G}^{-1} X (P^2=M^2_B)=0$,
where $X$ is the baryon wavefunction. Explicitly,
\be (1 - {\cal G} Z) X (P^2=M^2_B)= 0 \ , \ee
at $P^2=M_B^2$, which is a simplified version of the more complete Eq.~(\ref{eq:fadeev}).

The technical simplification of the Fadeev equation is given in App.~\ref{app:fadeev}. There, we apply
the ``static approximation'' for the exchanged
quark~\cite{Buck:1992wz}. This amounts to neglecting the exchanged momentum in $Z$ with respect to the quark
mass $m_l$. This approximation has been intensively used in other works resulting a very good
approximation (the estimated uncertainty is around 5\% as quoted in~\cite{Buck:1992wz}).

  As shown in App.~\ref{app:fadeev} the Fadeev equation can be recasted into a Dirac-like equation for the baryon
wavefunction evaluated at the baryon mass. For a particular baryon channel $BB'$ (we denote by $B,B'$ the 
physical baryon in the initial and final state), the equation to be solved reads
\be \label{eq:Dirac} \left. \left[ g^{\alpha \beta} \delta^{BB'} - M^{BB', \alpha \beta} (P^2) \right] \right|_{P^2=m_B^2}=0 \ , \ee
with the matrix (in both flavor and spin spaces) $M^{BB',\alpha \beta}$ introduced in Eq.~(\ref{eq:diracM}),
\begin{widetext}
\be M^{BB',\alpha \beta} (P) \equiv \frac{2}{m_l} \int \frac{d^4 q}{(2\pi)^4} \
 {\cal P}^{\dag,B}_{j \bar{j}} {\cal P}^{B'}_{\bar{k} k} \ \ 
 \tau^{\bar{k}}_{jl} \ \tau^{\bar{j}}_{lk} \ \  \Gamma^{\mu} \Gamma_{\mu}
 \ \ S_k \left( \frac{P}{2}+q \right) \ it^{\alpha \beta}_{\bar{k}} \left( \frac{P}{2}-q \right)  \ . \ee
\end{widetext}
For the members of the baryon octet, $M^{BB'}$ simplifies to (\ref{eq:Moctetpre})
\be \label{eq:Moctet} M^{BB'} (P) = \frac{2}{m_l} \ {\cal P}^{\dag,B}_{j\bar{j}} {\cal P}^{B'}_{\bar{k} k} \
\tau_{jl}^{\bar{k}} \tau_{lk}^{\bar{j}}  \  \Pi_{k \bar{k}} (P) \ , \ee
where the flavor matrices $\tau$ are given on the top of Table~\ref{tab:sextet}, the projection matrices ${\cal P}$
are given in App.~\ref{app:projections}, and the quark-diquark polarization
function is defined as
\be \label{eq:qdiqpol} \Pi_{k \bar{k}} (P) \equiv -\int \frac{d^4q}{(2\pi)^4} S_k (P-q) \ it_{\bar{k}} (q)  \ . \ee 

For the baryon decuplet the matrix $M^{BB',\alpha \beta}$ reads (\ref{eq:Mdecupletpre})
\be \label{eq:Mdecuplet} M^{BB',\alpha \beta} = \frac{8}{m_l} \ {\cal P}^{\dag,B}_{j\bar{j}} {\cal P}^{B'}_{\bar{k} k}
 \ \tau_{jl}^{\bar{k}} \tau_{lk}^{\bar{j}}
 \  \Pi^{\alpha \beta}_{k \bar{k}} (P) \ , \ee
where the flavor matrices for axial diquarks are given in the bottom of Table~\ref{tab:sextet} and the
projection matrices are also given in App.~\ref{app:projections}. The quark-diquark polarization function
is defined as
\be \label{eq:q-diq} \Pi^{\alpha \beta}_{k \bar{k}} (P) \equiv \int \frac{d^4q}{(2\pi)^4} S_k (P-q) \ it^{\alpha \beta}_{\bar{k}} (q)  \ , \ee 
where its final expression is given in App.~\ref{app:barpol}.

    In summary, the baryon masses are obtained by solving Eq.~(\ref{eq:Dirac}) with the matrix $M^{BB'}$
defined in~(\ref{eq:Moctet}) for the members of the baryon octet and in~(\ref{eq:Mdecuplet}) for the members of the
decuplet. The results for the masses and the extraction of the melting temperature for the different states are given
in the next section.

\section{\label{sec:conclusions} Results and Conclusions}

\begin{figure}[htp]
\begin{center}
\includegraphics[scale=0.42]{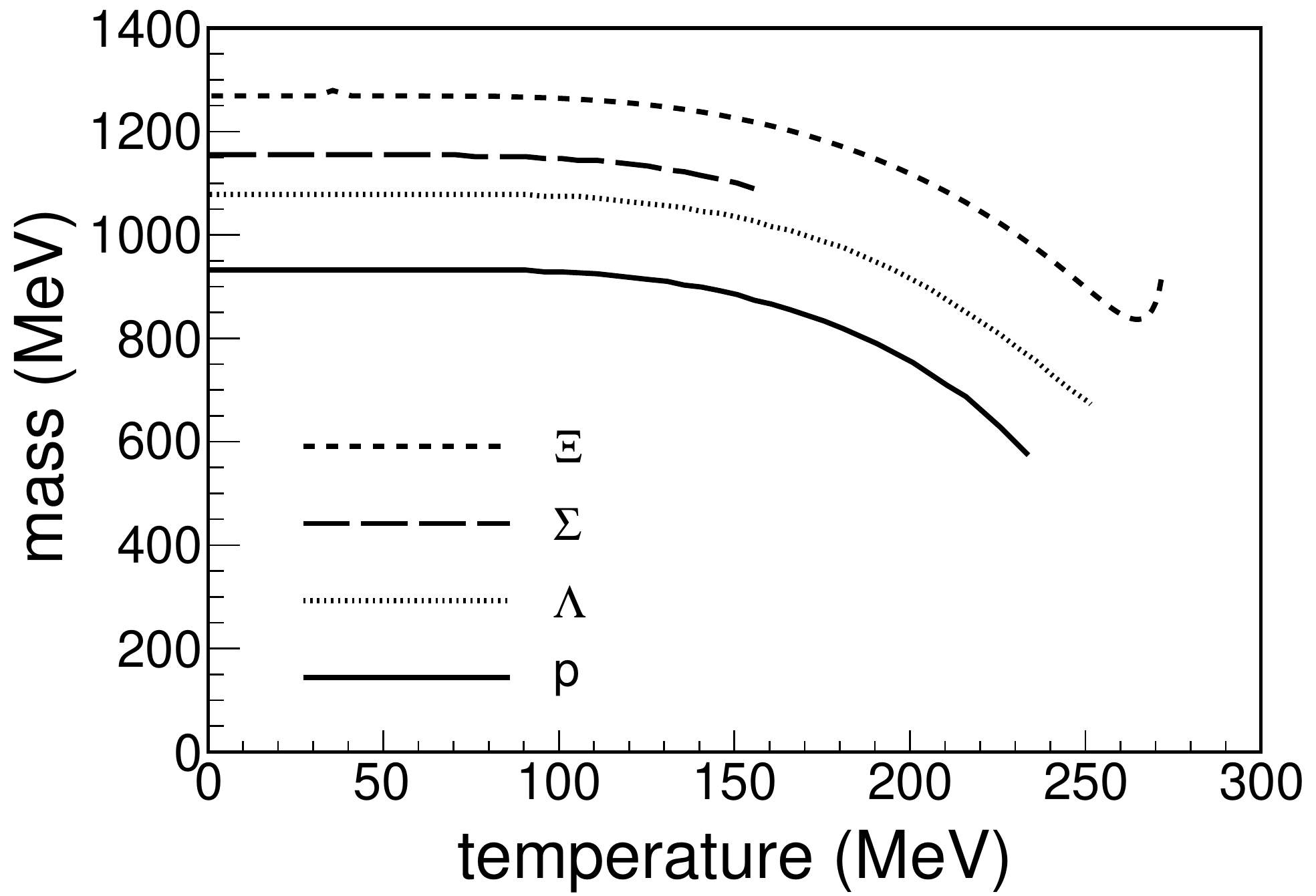}
\includegraphics[scale=0.42]{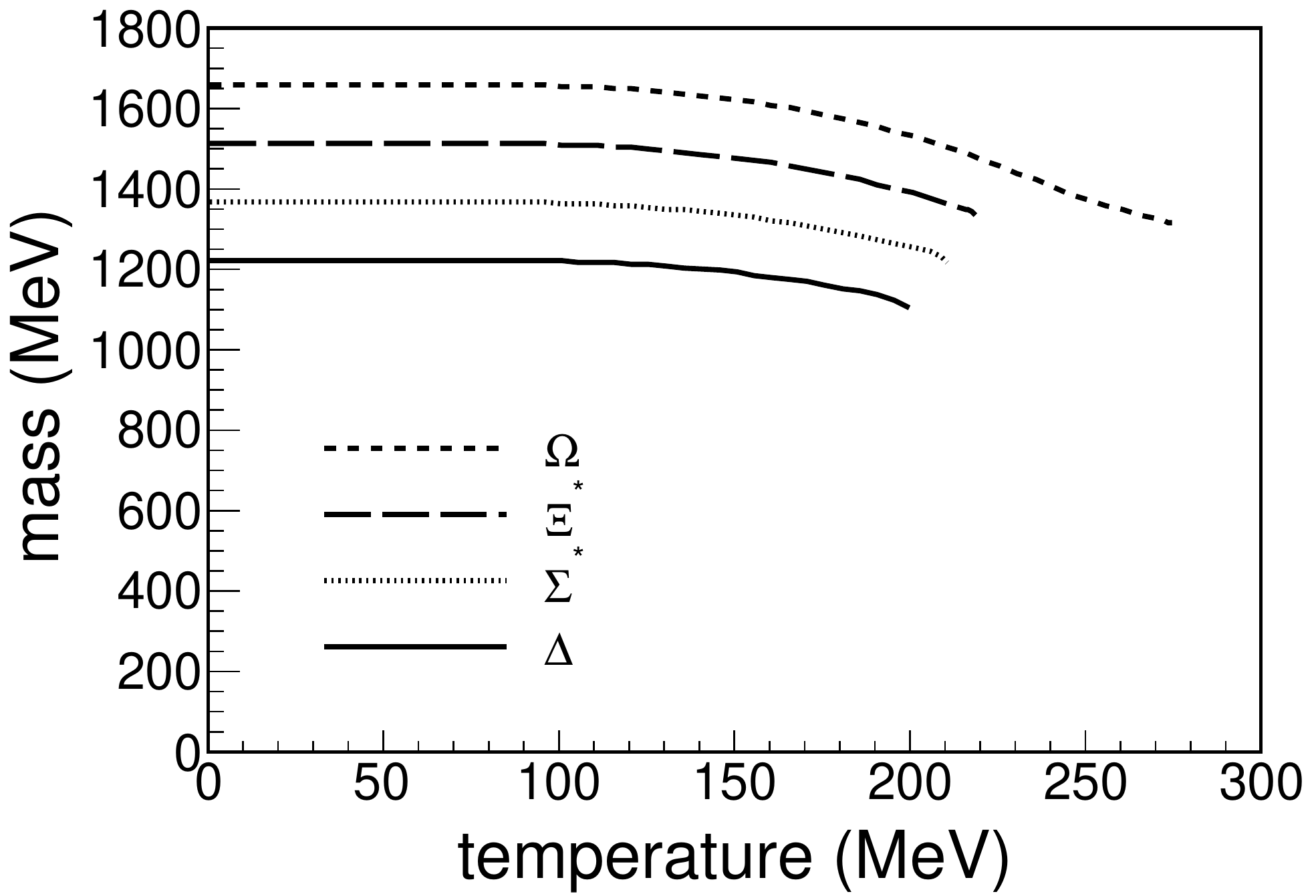}
\end{center} 
\caption{\label{fig:bar_njl} Baryon masses as a function of temperature for vanishing chemical potential
in the NJL model.}
\end{figure}

\begin{figure}[htp]
\begin{center}
\includegraphics[scale=0.42]{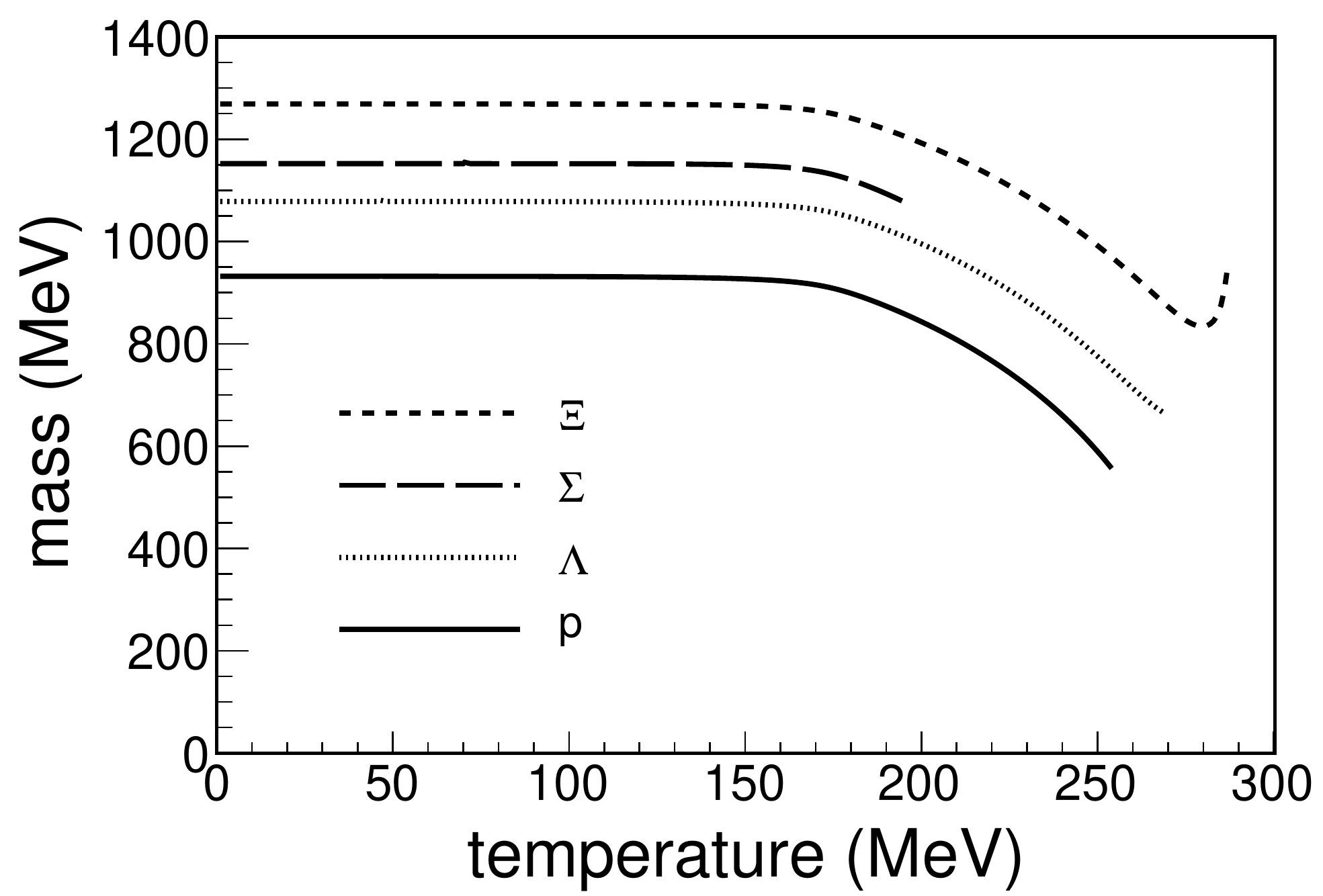}
\includegraphics[scale=0.42]{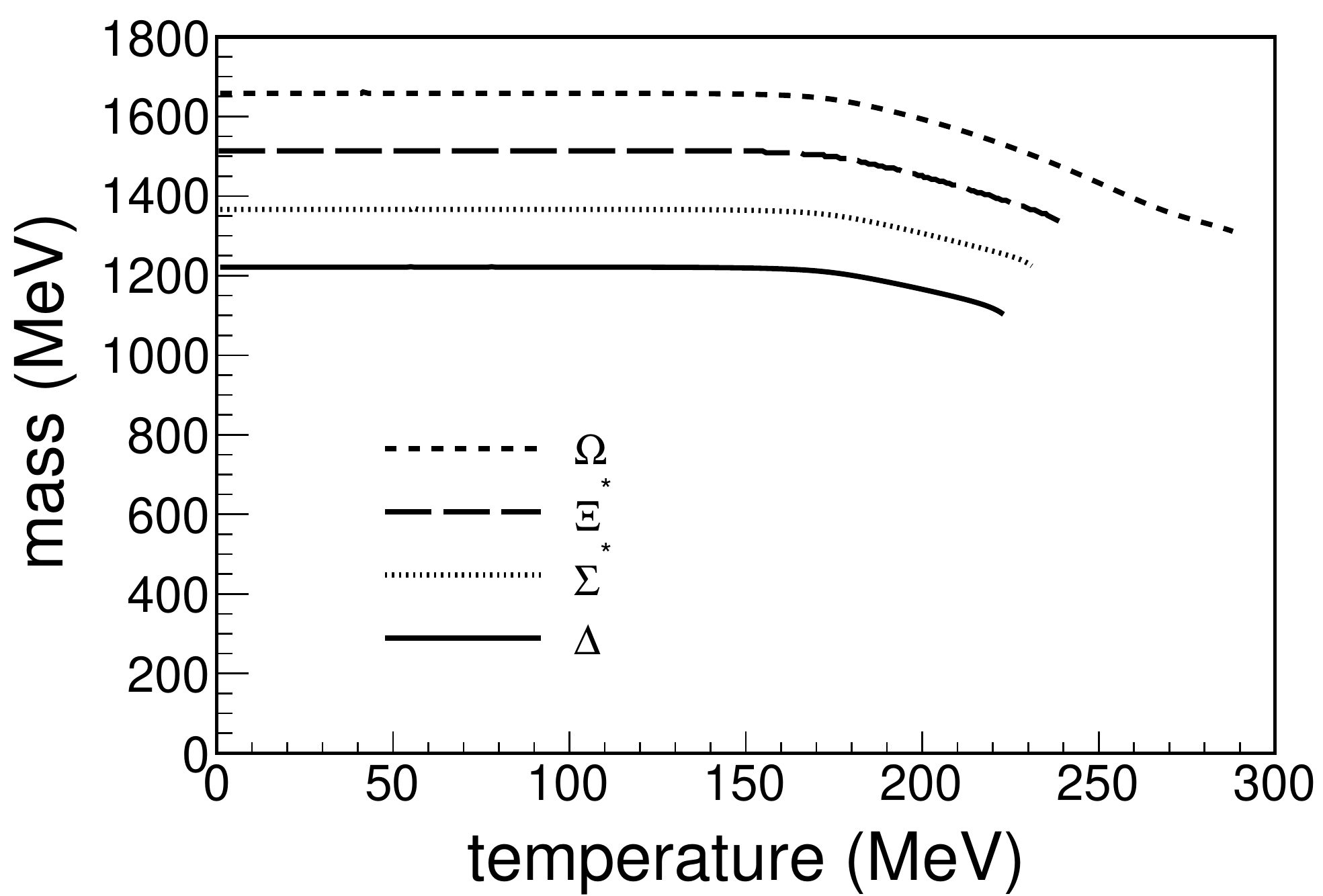}
\end{center} 
\caption{\label{fig:bar_pnjl} Baryon masses as a function of temperature for vanishing chemical potential
in the PNJL model.}
\end{figure}

   Our results for the baryon masses at finite temperature in both the octet and decuplet representations are shown in
Fig.~\ref{fig:bar_njl} for the NJL model. The results using the PNJL model are presented in Fig.~\ref{fig:bar_pnjl}.
We summarized all the baryon masses in vacuum ($T=0$) in Table~\ref{tab:barmass}.

From this example, the axial diquark case is straightforward. The only differences in the polarization function are
the numerical prefactor and the explicit appearance of Dirac indices in (\ref{eq:qdiqpol}). In addition, note
that the difference in the global sign cancels with the different sign in the diquark propagators
[cf. Eqs.~(\ref{eq:propscalar}] and (\ref{eq:propaxial})). Finally, we have neglected the term proportional to $p^\mu p^\nu$
in Eq.~(\ref{eq:propaxial}).

  The masses are given as a function of temperature for vanishing chemical potential.
As baryons are considered as quark-diquark bound states, the definition of the deconfinement temperature is
slightly modified to take into account the possibility of diquark melting. The ``deconfinement'' temperature may
be a genuine Mott temperature $T_{Mott}$(baryon), i.e., when the baryon mass equals the sum of quark and diquark masses.
Nevertheless, it can also happen that the diquark melts at a lower temperature $T_{Mott} (\textrm{diquark})
< T_{Mott} (\textrm{baryon})$. Therefore, we define the baryon deconfinement temperature as
the minimum of the two:
\be T_d \equiv \min \{T_{Mott} (\textrm{baryon}), T_{Mott} (\textrm{diquark}) \} \ , \ee
excluding the possibility that the three-quark system is stable when the diquark becomes unstable.

  Baryon masses are plotted up to this temperature, which is summarized
in Table~\ref{tab:barmass} for all the baryon states. Beyond $T_d$ the baryon mass becomes complex
and, in principle, one should perform an analytical continuation of the Matsubara frequencies to nonreal
energies. This is beyond the scope of this work.

\begin{table}
\begin{center}
\begin{tabular}{|c|c|c|c|c|c|c|c|c|}
\hline
Baryon & $p$ &$\Lambda$&$\Sigma$&$\Xi$&$\Delta$&$\Sigma^*$&$\Xi^*$ &$\Omega$\\
\hline
\hline
Mass at $T=0$  & 932 & 1078 & 1152 & 1269&1221&1366&1512&1658 \\
Mass - pdg~\cite{Beringer:1900zz}  & 939&1116&1193&1318&1232&1383&1533&1672 \\
\hline
$T_{Mott}$ - NJL & 234 & 252 & 156& 272 & 200 & 211 & 219 & 275 \\
$T_{Mott}$ - PNJL & 254 & 269 & 195 & 287 & 223 & 231 & 239 & 288 \\
\hline
\end{tabular}
\caption{\label{tab:barmass} Masses at $T=0$ and Mott temperature for the different baryons in the octet and decuplet representations
for both NJL and PNJL models. All units are given in MeV.}
\end{center}
\end{table}  

    Comparison with previous results~\cite{Gastineau:2002jh,Gastineau,Blanquier,Blanquier:2011zz} show a
similar trend for all the masses. In particular the surprisingly low melting temperature for $\Sigma$ baryon
is also captured by~\cite{Gastineau:2002jh,Gastineau} where, in addition, the melting temperature
of proton is slightly larger than the $\Lambda$, as opposed to our case. In Refs.~\cite{Blanquier,Blanquier:2011zz} the
decuplet is shown for the first time. However, in these works the Dirac structure of the Dirac equation
(\ref{eq:diracM}) is simplified by taking the trace in the quark-diquark polarization function (with no further 
justification for this step). In addition, the transverse nature of the axial diquark propagator is omitted.
In spite of this fact, the trend for the baryon octet coincides with our results, also in the larger mass
of the $\Lambda$ with respect to the proton. The decuplet sector is also similar to ours.

In the decuplet sector we find a hierarchy based on the strangeness content, i.e. the $T_c$ increases
with the number of constituent strange quarks of the baryon. In this respect the first baryon that melts is the $\Delta$, whereas 
the $\Omega$ has the largest Mott temperature.

    The main conclusion of this work is the evident flavor dependence of the deconfinement temperature of baryons.
On the qualitative level, our result is quite robust due to the fact that the relative temperatures remain similar
for both NJL and PNJL models. On the quantitatively side, the Mott temperatures are strongly dependent on the
model (NJL or PNJL) used, and in both cases they seem to overestimate the standard values for the freeze-out temperatures obtained by the thermal fits.

    Statistical models applied to ALICE data predict a lower chemical freeze-out temperature for proton in
comparison with the that for states with multistrangeness ($\Xi$ and $\Omega$). Within our approximation, our findings
agree with this claim being the proton temperature 38 MeV (33 MeV) less than the $\Xi$ temperature in the NJL
(PNJL) model. This fits well to the experimental findings in~\cite{Preghenella:2011np}.

    In addition, we find that the temperature of the $\Xi$ and $\Omega$ baryons are surprisingly similar, in
accordance with the thermal-statistical model result~\cite{Preghenella:2011np}. In our scheme, this result is
totally nontrivial, because of the fact that the internal baryon structure is entirely different for the two states, because we have
different kinds of constituent diquarks.

    In summary, we have applied the NJL and PNJL models, together with different many-body techniques ---Bethe-Salpeter 
and Fadeev equations--- to generate diquarks and baryons, respectively. The parameters of the models are fitted to
agree with the low-lying states in the respective channel ($G_{DIQ}$ fitted to reproduce the proton mass and
$G_{DIQ,V}$ to the $\Delta$ baryon mass), being the mass of other baryons predictions of the models.

    Extending the method to finite temperature we were able to predict the temperature dependence of baryon masses
for all the physical states belonging to the flavor octet and decuplet representations. We find a strong 
dependence on the melting (or deconfinement) temperature depending on the flavor content of the baryons. In a
qualitative way, our findings coincide with the suggested results on the statistical thermal model on ALICE
data~\cite{Preghenella:2011np}, and supported by lattice-QCD results by~\cite{Bellwied:2013cta}.

\acknowledgments

We acknowledge M. Buballa and L. Tol\'os for interesting and very useful discussions. This research has been supported by the
Program TOGETHER from R\'egion Pays de la Loire and the
European I3-Hadron Physics program. J.M.T.-R. has also been funded by Ministerio de
Ciencia e Innovaci\'on under Contracts No. FPA2010-16963 and No. FPA2013-43425-P.

\appendix

\section{Fierz transformation\label{app:fierz}}

   The Fierz transformation allows us to convert the original NJL Lagrangian in Eq.~(\ref{eq:lagNJL}) ---based on
the color-current interaction--- into a Lagrangian where the fermion fields are reordered to account for
interactions in different color, flavor and spin sectors.

   Even if the Fierz transformation is a well-known tool~\cite{Fierz}, some differences in its application can
be found in the context of the NJL model. The Fierz transformation itself is based on algebraic
identities in flavor, color and spin spaces. It is therefore unique once $N_f$ and $N_c$ are fixed.
However, there are at least two methods to obtain the transformed Lagrangian.

  The first method requires two different Fierz transformations~\cite{Buballa:2003qv} to obtain all the mesonic
and diquark contributions. A Fierz transformation transforms the original color-current interaction to the
``exchange sector'' containing the $\bar{q}q$ interaction in both color singlet and
color octet representations. In color space the transformation reads:
\be \sum_{a'=1}^{N_c^2-1} T^{a'}_{i'j'} T^{a'}_{k'l'} = 2 \frac{N_c^2-1}{N_c^2} \delta_{i'l'} \delta_{k'j'} - \frac{1}{N_c} \sum_{a'=1}^{N_c^2-1}
T^{a'}_{i'l'} T^{a'}_{k'j'} \ . \ee

  To study physical mesons, one considers the first term, while the second term is simply neglected. A second Fierz transformation is applied to
the color-current interaction to generate the $qq$ sector in both color antitriplet and sextet representations:
\be \sum_{a'=1}^{N^2_c-1} T^{a'}_{i'j'} T^{a'}_{k'l'} = \frac{N_c-1}{N_c} \sum_{S'} T^{S'}_{i'k'} T^{S'}_{l'j'} - \frac{N_c+1}{N_c}
\sum_{A'} T^{A'}_{i'k'} T^{A'}_{l'j'} \ , \ee
where $S'$ and $A'$ run over the symmetric and antisymmetric members of the color representation, respectively.
For $N_c=3$, $S'=1,3,4,6,8$ and $A'=2,5,7$. The first term is neglected as it contains a repulsive
diquark interaction and is not useful to generate colorless baryons.

  The second method makes use of only one Fierz transformation~\cite{Vogl:1991qt,Alkofer:1995mv} which follows 
from the principle of obtaining only attractive-color interactions. With this prescription, the
final Lagrangian for both mesons and diquarks is obtained by a single Fierz transformation in color space:
\be \label{eq:2fierz} \sum_{a'=1}^{N_c^2-1} T^{a'}_{i'j'} T^{a'}_{k'l'} = \frac{N_c-1}{2N_c} \delta_{i'l'} \delta_{j'k'}
- \frac{2}{N_c} \sum_{A'} T_{i'k'}^{A'} T_{l'j'}^{A'} \ .  \ee
 In the right-hand side of Eq.~(\ref{eq:2fierz}), the first term produces the Lagrangian for mesons in the color singlet representation, and the second one the
Lagrangian for diquarks in the color antitriplet-color representation. No sign of any $\bar{q}q$ interaction in
the color octet and of the $qq$ one in the color sextet is seen. As claimed, these channels are not needed to obtain
physical mesons and the baryonic states.
   
 Notice that the numerical factors appearing in front of the Fierzed Lagrangian depend on the choice of the 
method. Therefore, they present different values for the coupling constants. However, we remind that we are
treating these couplings as free parameters to be fixed by reproducing the hadron masses. For this reason
the two methods are equivalent, if the coupling constant is considered as a free parameter.

  In this work we will use the conventions given in Ref.~\cite{Buballa:2003qv} and generate all possible interactions
in the $qq$ and in the $\bar{q}q$ sectors. The terms which we obtain are in exact correspondence with the
irreducible representations of the flavor and color group products.

  For the meson sector the Fierz-transformed Lagrangian reads~\cite{Buballa:2003qv} (we suppress the flavor
indices $i,j$):
\begin{widetext}
\be \label{eq:Lex}
\begin{array}{ccccccccc}
  {\cal L}_{ex} &=& \frac{2(N_c^2-1)}{N_f N_c^2} g  \left[ \right. (\bar \psi \psi)^2  &+& (\bar \psi i\gamma_5  \psi)^2 
&-& \frac{1}{2}(\bar \psi \gamma^\mu  \psi)^2 &-& \frac{1}{2}(\bar \psi \gamma^\mu\gamma_5  \psi)^2 \left.\right] \\
  &+& {\bf \frac{N_c^2-1}{N_c^2} g  \left[ \right.  (\bar \psi \tau^a \psi)^2} &{\bf +}& {\bf (\bar \psi i\gamma_5 \tau^a  \psi)^2 }
&{\bf -}& {\bf \frac{1}{2}(\bar \psi \gamma^\mu \tau^a \psi)^2 }& {\bf -}& {\bf \frac{1}{2}(\bar \psi \gamma^\mu\gamma_5 \tau^a \psi)^2 \left. \right]}  \\
&-& \frac{1}{N_fN_c} g \left[ \right. (\bar \psi T^{a'} \psi)^2 &+& (\bar \psi i\gamma_5T^{a'}  \psi)^2 
&-& \frac{1}{2}(\bar \psi \gamma^\mu T^{a'}  \psi)^2 &-& \frac{1}{2}(\bar \psi \gamma^\mu\gamma_5T^{a'}  \psi)^2 \left. \right] \\
&-& \frac{1}{2N_c} g \left[ \right. (\bar \psi\tau^aT^{a'} \psi)^2 &+& (\bar \psi i\gamma_5\tau^aT^{a'}  \psi)^2
& -& \frac{1}{2}(\bar \psi \gamma^\mu\tau^aT^{a'}  \psi)^2 &-& \frac{1}{2}(\bar \psi \gamma^\mu\gamma_5\tau^aT^{a'}  \psi )^2 \left. \right] \ , 
\end{array}
\ee
\end{widetext}
where $a=1,...,N_f^2-1$ and $a'=1,...,N_c^2-1$. It is not difficult to check that the different terms match with 
the representations spanned by
\be ({\bf 3} \otimes {\bf \bar{3}})_c \otimes ({\bf 3} \otimes {\bf \bar{3}})_f \otimes
 \{ 1, i\gamma_5,\gamma_\mu, \gamma_5 \gamma_\mu \} \ . \ee
The second row (in bold font) will be of interest to produce the physical mesons.
In particular, the second term is the relevant one for the pseudoscalar mesons ($\pi,K$ and $\eta$).

  Performing the second Fierz transformation one obtains the diquark sector~\cite{Buballa:2003qv}:

\begin{widetext}
 \ba
 \begin{array}{cccccc}
{\cal L}_{qq} & = & \frac{N_c+1}{2N_c} g & \left[ \right. {\bf (\bar \psi i\gamma_5C \tau^AT^{A'}\bar \psi^T)
       (\psi^T C i\gamma_5 \tau^AT^{A'} \psi) } & +&  (\bar \psi C \tau^AT^{A'}\bar \psi^T)
       (\psi^T C  \tau^AT^{A'} \psi) \\
 & - & & \frac{1}{2} (\bar \psi \gamma^\mu\gamma_5C \tau^AT^{A'}\bar \psi^T)
       (\psi^T C \gamma_\mu\gamma_5 \tau^AT^{A'} \psi) &\bf{-} & \bf{ \frac{1}{2} 
       (\bar \psi\gamma^\mu C \tau^ST^{A'}\bar \psi^T)
       (\psi^T C\gamma_\mu  \tau^ST^{A'} \psi)} \left. \right] \\
 & - & \frac{N_c-1}{2N_c} g & \left[ \right. (\bar \psi i\gamma_5C \tau^ST^{S'}\bar \psi^T)
       (\psi^T C i\gamma_5 \tau^ST^{S'} \psi) &+&  
       (\bar \psi C \tau^ST^{S'}\bar \psi^T) (\psi^T C  \tau^ST^{S'} \psi) \nonumber\\
& -& &\frac{1}{2} (\bar \psi \gamma^\mu\gamma_5C \tau^ST^{S'}\bar \psi^T)
       (\psi^T C \gamma_\mu\gamma_5 \tau^ST^{S'} \psi) & -&\frac{1}{2} 
       (\bar \psi\gamma^\mu C \tau^AT^{S'}\bar \psi^T)
       (\psi^T C\gamma_\mu  \tau^AT^{S'} \psi) \left. \right] \ ,  
\end{array}  \\  \label{eq:Lqq}
\ea
\end{widetext} 
where the indices $A,A'$ refer to the antisymmetric members of the flavor and color representations, respectively, and
the indices $S,S'$ to the symmetric elements of the flavor and color representations. In this way, the
Fierz transformation in the $qq$ sector generates the eight possible terms consistent with the direct product of
\be {\cal A} \ [ \ ({\bf 3} \otimes {\bf 3})_c \otimes ({\bf 3} \otimes {\bf 3})_f \otimes 
\{ 1, i\gamma_5,\gamma_\mu, \gamma_5 \gamma_\mu \} \ ] \ , \ee
where ${\cal A}$ denotes the antisymmetrization operator to respect the  Pauli principle for the exchange of two quarks.
In Eq.~(\ref{eq:Lqq}) we highlight the two terms giving rise to the scalar and axial diquark discussed in this work when forming baryons.

\section{Quark condensate and meson polarization function~\label{app:AB}}

  For completeness, we include here some reduced expressions for the quark condensate 
in Eq.~(\ref{eq:conden}) and the meson polarization function in Eq.~(\ref{eq:polmeson}). For practical reasons we 
remind here the $A$ and $B_0$ functions introduced in Ref.~\cite{Rehberg:1995kh,Rehberg:1995nr}.
These functions are convenient for implementing numerically the momentum integrations for the thermal averages.

For vanishing chemical potentials, the function $A$ is defined as
\be A (m_i,T,\Lambda) \equiv 16 \pi^2 T \sum_n \int \frac{d^3k}{(2\pi)^3} \frac{1}{ (i\omega_n)^2 - E_i^2} \ , \ee
with the quark energy $E_i=\sqrt{k^2+m_i^2}$ and the fermionic Matsubara frequencies
$i\omega_n = i (2n+1) \pi T$. This function naturally appears when computing the quark condensate defined
in Eq.~(\ref{eq:conden}):
\be \label{eq:condennjl}\langle \bar{\psi}_i \psi_i \rangle = N_c \frac{m_i}{4\pi^2} A(m_i,T,\Lambda) \ . \ee
Performing the Matsubara summation~\cite{Kapusta:2006pm} one obtains
\ba A (m_i,T, \Lambda) &=& -16 \pi^2 \int \frac{d^3k}{(2\pi)^3} \frac{1}{2E_i} \left[ 1 - 2n_F (E_i) \right]  \\
&=& 4 \int_{m_i}^{\Lambda_E} dE_i \sqrt{E_i^2-m_i^2} \left[ n_F(E_i) - n_F (-E_i) \right] \ , \nn \ea
where $n_F(E_i)=(e^{E_i/T}+1)^{-1}$ is the Fermi-Dirac distribution function and $\Lambda_E=\sqrt{\Lambda^2+m_i^2}$.
For the PNJL model one simply replaces Eq.~\ref{eq:condennjl} by Eq.~\ref{eq:condenpnjl}.

The function $B_0$ is defined as
\begin{widetext}
 \be B_0 (m_i,m_j, i \nu_m, {\bf p}, T, \Lambda) = 16 \pi^2 T \sum_n \int \frac{d^3k}{(2\pi)^3} \frac{1}{(i\omega_n)^2 - E_i^2}
\frac{1}{(i\omega_n - i\nu_m)-E_j^2}\ , \ee
\end{widetext}
with $E_i=\sqrt{k^2+m_i^2}$ and $E_j=\sqrt{({\bf k} - {\bf p})^2+m_j^2}$. It naturally appears in the quark-(anti)quark loop function, when computing the diquark (meson) polarization function.
After performing the Matsubara summation, the Matsubara frequency $i\nu_m$ is analytically continued to real values of the energy with the prescription $i\nu_m \rightarrow p_0 + i \epsilon$
For instance, the pion polarization function ($m_i=m_u$, $m_j=m_d$) can be expressed as a combination of the $A$ and $B_0$ functions as
\begin{widetext}
\ba  \Pi^{ud} (p_0,{\bf p},m_u,m_d,T,\Lambda) = -\frac{N_c}{4\pi^2} \left\{ A(m_u,T,\Lambda) + A(m_d,T,\Lambda)+
 \left[ (m_u-m_d)^2-p_0^2+{\bf p}^2 \right] B_0 (m_u,m_d,p_0,{\bf p},T,\Lambda) \right\} \nn \  . \\ \label{eq:pipion} \ea
\end{widetext}
In the case of a pion at
rest, ${\bf p}=0$, one can perform the Matsubara summation and finds 
\begin{widetext} 
\be  B_0 (m_i,m_j, p_0, 0, T, \Lambda) = \sum_{\sigma=\pm} 
\ \sigma \left[ B_0^\sigma (m_i,m_j,-\sigma p_0,T,\Lambda)  + B_0^\sigma (m_j,m_i,\sigma p_0,T,\Lambda) 
\right] , \ee
where 
\ba \Re \ B_0^{\pm}(m_i,m_j,p_0,T,\Lambda) &=& \frac{2}{p_0} \textrm{ P.V.} \int_{m_i}^{\Lambda_E} dE_i \sqrt{E_i^2-m_i^2} \frac{n_F(\pm E_i)}{E_i-E_0} \ , \\
\label{eq:imB0} \Im \ B_0^{\pm}(m_i,m_j,p_0,T,\Lambda) &=& \frac{2\pi}{p_0} \sqrt{E_0^2-m_i^2} \ n_F(\pm E_0) \Theta(\Lambda_E-E_0) \Theta(E_0-m_i) \ , 
\ea
\end{widetext}
with $E_0 \equiv - (p_0^2+m_i^2-m_j^2)/(2p_0)$. \\

 Let us briefly discuss some of the prescriptions to define the mass and decay width of the mesons and diquarks.

 These generated states are identified with the poles of the scattering amplitude $t^{ab}(p^2)$, or
Eq.~(\ref{eq:mesonmass}). We can distinguish those states that are generated below and above the two-quark
mass threshold, that is, those which cannot decay into a pair of quarks and those in which this decay channel is
open because their mass is larger than the combined mass of the constituents.

 In the first case we talk of ``bound states'' (the decay width is exactly zero). For them, the polarization
function is a real function ($B_0$ does not develop any imaginary part) and the
pole is generated on the real axis of the $p_0$-plane (in the first Riemann sheet). The real value of $p_0$ is
associated with the mass of the bound state. In the second case we denote them as ``resonances'' (a finite decay
width is generated), the polarization function is now complex [Eq.~(\ref{eq:imB0}] is nonzero) and the pole emerges at a complex $p_0$ (in the
second Riemann sheet). The imaginary part of the pole can be related to the decay width of the resonance.

 At finite temperature, the variable $p_0=i\nu_m$ is strictly a Matsubara frequency. If a bound state is generated, then
one can simply make the standard analytical continuation to real energies $i\nu_m \rightarrow p_0 + i \epsilon$ and
find the value of the generated mass by solving $1-2K^{ab} \Pi^{ab} (p_0)=0$, with real $\Pi^{ab} (p_0)$. However, for
a resonant state, one must analytically continue the Matsubara frequency to complex energies and find the pole in the
second Riemann sheet. As this procedure might be cumbersome (in particular for cases with several coupled channels, 
where additional Riemann sheets must be considered), approximate methods are used (see, for instance,~\cite{Hansen:2006ee}). 

  One introduces the spectral density $\rho(p_0,{\bf p})$ (see~\cite{Mahan}, for instance) as the imaginary part of the 
bound state/resonance propagator,
\ba \rho^{ab}(p_0,{\bf p}) &=& -\frac{1}{\pi} \Im t^{ab} (p_0,{\bf p}) \\
&=&  \frac{1}{\pi} \frac{\Im 
\Pi^{ab} (p_0,{\bf p})}{[ (2G)^{-1} - \Pi^{ab} (p_0,{\bf p}))^2+(\Im \Pi^{ab} (p_0,{\bf p})]^2} \nn \ , \ea
where the Matsubara frequencies has been analytically continued to real energies. Therefore, the spectral function 
is a real function of real argument. 

  Note that taking ${\bf p}=0$, using the pole approximation and the optical theorem we can check that the
spectral function is proportional to the scattering amplitude squared evaluated on the real axis,
\be \rho^{ab} (p_0,0) = -\frac{1}{\pi} \
\Im t^{ab} (p_0,0)  \propto |T|^2 (p_0,0) \ . \ee 
  Thus, whenever the scattering amplitude presents a pole, this is reflected into the spectral function as a
peak. If the pole is not far from the real axis ($\Im p_0 \ll \Re p_0$), the real part of $p_0$ at the pole position 
coincides with the maximum of the spectral function. Therefore, the mass can be defined as the position of the
spectral density maximum. If, in addition, $\Im \Pi^{ab} (p_0)$ is a smooth function of $p_0$~\cite{Mahan} around
the peak, it can be approximated by a Lorentzian shape, with a width~\cite{Mahan,Hansen:2006ee}
\be \Gamma = - \Im \Pi^{ab} (p_0) \ . \ee
In this approximation, one defines the decay width of the resonance as the Lorentzian width $\Gamma$.

  However, this approximation might break down if the pole position is far away from the real axis, and in particular,
if the many-body equation presents coupled channels: a broad resonance can be hidden by another pole with more strength
in this channel, several poles might appear very close in the complex plane producing a combined shape of the
the spectral function, a new threshold opens close to the resonance (Flatt\'e effect) blurring the Lorentzian shape, etc. 

  As in this paper we are not interested in a precise extraction of the decay widths but only in the temperature 
at which they become nonzero, we use an intermediate prescription described
in Ref.~\cite{Rehberg:1995kh,Rehberg:1995nr}. In this case one makes the analytical continuation to real
energies (after having performed the Matsubara summation) but considers a complex $p_0$ in the factor in 
front of the $B_0$ function in Eq.~(\ref{eq:pipion}). This prescription provides a complex polarization function
of complex argument and one has direct access to the mass ($m=\Re p_0$) and decay width 
($\Gamma=- 2 \Im p_0$) of the resonant state.

\section{Bethe-Salpeter equation for quark-quark scattering\label{app:bs}}

The Bethe-Salpeter equation for the $qq$-scattering~\cite{Lehmann,Mu:2012zz,Sun:2007fc} in the scalar channel reads Eq.~(\ref{eq:BSdiquark})
\begin{widetext}
\be T^{ab}_{ij,mn} (p^2) = {\cal K}^{ab}_{ij,mn} + i \int \frac{d^4 k}{(2 \pi)^4}
{\cal K}^{ac}_{ij,pq} \  S_{p} \left( k+ \frac{p}{2} \right) \ S^c_{q}  \left( \frac{p}{2}-k \right) \ T^{cb}_{pq,mn} (p^2) \ . \ee
\end{widetext}

We can pull out all the vertex factors by defining the diquark propagator $t_{ab}$
\be T^{ab}_{ij,mn} (p^2) = \Omega^a_{ij} t^{ab} (p^2) \bar{\Omega}^b_{nm} \ , \ee
where
\be \Omega^a_{ij} = T^{a'} \otimes \tau_{ij}^a \otimes \Gamma C \ , \ee
with $\Gamma=i \gamma_5\gamma_\mu$ for scalar and axial diquarks, respectively; $C=i\gamma_0 \gamma_2$ being the charge conjugation operator.

The kernel is taken from the Fierzed NJL Lagrangian of Eq.~(\ref{eq:lagdiq})
\be {\cal K}^{ab}_{ij,mn} = \Omega^a_{ij} \ 2 G_{DIQ} \ \bar{\Omega}^b_{nm} \ , \ee
where the 2 is a combinatorial factor arising when attaching the external legs to the vertex and $G_{DIQ}$ must be substituted by $G_{DIQ,V}$ for axial diquarks. 
Note that we neglect any contribution from the 't Hooft Lagrangian as there is no flavor singlet in this channel and its effects are expected to be much suppressed (there is a small effect around
4 \%, discussed in Ref.~\cite{Lehmann}).

In terms of the amplitude $t(p^2)$ we can express the solution of the BS equation for scalar diquarks,
\be \label{eq:ampdiq} t^{ab}(p^2) = \frac{2 G_{DIQ} }{1- 2  G_{DIQ} \Pi^{ab} (p^2) } \ , \ee
with the polarization function 
\be \label{eq:poldiq} \Pi^{ab} (p^2) = i \int \frac{d^4 k}{(2\pi)^4} \textrm{ Tr }
 \left[ \bar{\Omega}^a_{ji}  S_i (k+p/2) \Omega^b_{ij} S^T_j (p/2-k)  \right] \ , \ee
where the trace is to be taken in color, flavor, and Dirac spaces. For axial diquarks, the amplitude involves the transverse part of the polarization function [cf. Eq.~(\ref{eq:components})],
\be \label{eq:ampdiq2} t^{ab}(p^2) = \frac{2 G_{DIQ,V} }{1- 2  G_{DIQ,V} \Pi_\perp^{ab} (p^2) } \ . \ee

For the scalar case one has
\begin{widetext}
\be \Pi^{ab} (p^2) =   \textrm{ tr}_c (T^{a'} T^{b'}) \
   \tau_{ji}^{a} \tau_{ij}^b \ i \int \frac{d^4 k}{(2\pi)^4} \textrm{ tr}_\gamma \left[ i\gamma_5 S_i
 \left( k + \frac{p}{2} \right)  i \gamma_5  C S^{T}_j \left( \frac{p}{2} -k\right)C^{-1} \right] \ . \ee 
\end{widetext}

The color factor is common to all diquarks in the antitriplet color representation, where the generators can be taken as~\cite{Lehmann}
\be \label{eq:color3bar} (T^{a'})_{j'k'} = i \epsilon_{a'j'k'} \ . \ee

The color factor reads
\be \textrm{ tr}_c (T^{a'} T^{b'}) =  -\sum_{k'l'} \epsilon^{a'k'l'} \epsilon^{b'l'k'} = 2 \delta^{a'b'} \ , \ee
which means that the color of the diquark does not change in the propagation. From now on we will suppress the 
color indices. 

Using the identity,
\be  C S^{T}_i \left( \frac{p}{2} -k\right) C^{-1}= S_i \left( k-\frac{p}{2}\right) \ , \ee
we can express the polarization function as
\be  \Pi^{ab} (p^2) =   2 \ \tau^{a}_{ji} \tau_{ij}^b \ i \int \frac{d^4 k}{(2\pi)^4} 
\textrm{ tr}_\gamma \left[ i\gamma_5 S_i \left( k  \right) i \gamma_5  S_j \left( k- p \right) \right] \ , \ee 
where we have performed a variable shift $k\rightarrow k - p/2$.

In flavor space, we choose the representations shown in Table~\ref{tab:sextet} for the sextet and antitriplet case~\cite{Oettel,Lehmann}.
For both of them, the normalization is $\textrm{tr } (\tau^a \tau^b)= 2 \delta^{ab}$.
\begin{table}
\begin{center}
\begin{tabular}{|c|c|}
\hline
Physical diquark & ${\bf \bar{3}}$ Representation  \\
\hline
\hline
$[ud]$ & $\tau^{[ud]}_{ij}=\tau^{\bar{3}}_{ij} =- \lambda^2_{ij} $ \\
$[us]$ & $\tau^{[us]}_{ij}=\tau^{\bar{2}}_{ij} = \lambda^5_{ij} $ \\
$[ds]$ & $\tau^{[ds]}_{ij}=\tau^{\bar{1}}_{ij} =- \lambda^7_{ij} $ \\
\hline
\hline
Physical diquark & ${\bf 6}$ Representation  \\
\hline
\hline
$(uu)$ & $\tau^{(uu)}_{ij}=\tau^{1}_{ij} = \sqrt{2} \delta_{i1} \delta_{j1} $ \\
$(ud)$ & $\tau^{(ud)}_{ij}= \tau^{2}_{ij}= \lambda^1_{ij} $ \\
$(dd)$ & $\tau^{(dd)}_{ij}=\tau^{3}_{ij}= \sqrt{2} \delta_{i2} \delta_{j2} $ \\
$(us)$ & $\tau^{(us)}_{ij}=\tau^{4}_{ij}= \lambda^4_{ij} $ \\
$(ds)$ & $\tau^{(ds)}_{ij}=\tau^{5}_{ij}= \lambda^6_{ij} $ \\
$(ss)$ & $\tau^{(ss)}_{ij}=\tau^{6}_{ij}=\sqrt{2}  \delta_{i3} \delta_{j3} $ \\
\hline
\end{tabular}
\caption{\label{tab:sextet} Flavor matrices for the antitriplet and sextet flavor representations of the
direct product ${\bf 3}_f \otimes {\bf 3}_f$ of $SU(3)$.}
\end{center}
\end{table}

As an example, we calculate the lightest $a=[ud]$ diquark. It is easy to see that the polarization
function is diagonal in flavor, so the only possibility is to have $b=[ud]$. Using $\tau^{[ud]}_{ij}=-\lambda^2_{ij}$,
\be \Pi^{[ud]} (p^2) =  4 i\int \frac{d^4k}{(2\pi)^4} \textrm{ tr}_\gamma \left[ i\gamma_5 S_d \left( k \right) i \gamma_5  S_u \left(k- p \right) \right] \ . \ee
This is the final expression for the polarization function that gives rise to the propagator of the $[ud]$ diquark.

At finite temperature, we introduce fermionic Matsubara frequencies,
\begin{widetext}
\be  \Pi^{[ud]} (i\nu_m,{\bf p}) =
 -4  T \sum_n \int \frac{d^3 k}{(2\pi)^3} \textrm{ tr}_\gamma \left[ i\gamma_5 S_d \left( i\omega_n, {\bf k}  \right) i \gamma_5  S_u \left( i\omega_n-i\nu_m,{\bf k- p} \right) \right] \ . \ee
\end{widetext}
After performing the Matsubara sum, we analytically continue the unsummed Matsubara frequency $i\nu_m$ to the real energy $p_0+i\epsilon$.
For practical purposes, this function can be reduced in terms of the $A, B_0$ functions defined in~\cite{Rehberg:1995kh,Rehberg:1995nr} and detailed in App.~\ref{app:AB}:
\begin{widetext}
\be  \Pi^{[ud]} (p_0,{\bf p}) =  -\frac{1}{2\pi^2} \left\{ A(m_u,T,\Lambda) + A(m_d,T,\Lambda) + \left[ (m_u-m_d)^2-p_0^2+{\bf p}^2\right] B_0 (m_u,m_d,p_0,{\bf p},T,\Lambda) \right\} \nn \  . \ee
\end{widetext}
Similar expressions can be found for other diquarks in different flavor channels.

For axial diquarks, the expression for the transverse part of the polarization function reads:
\begin{widetext}
\be  \Pi_\perp^{(ud)} (p_0,{\bf p}) =  \frac{1}{3\pi^2} \left\{ A(m_u,T,\Lambda) + A(m_d,T,\Lambda) + \left[ (m_u-m_d)^2-2m_um_d-p_0^2+{\bf p}^2\right] B_0 (m_u,m_d,p_0,{\bf p},T,\Lambda) \right\} \nn \  . \ee
\end{widetext}

\section{Reduction of the Fadeev equation \label{app:fadeev}}
  
  In this appendix we will give some details of the simplification of the Fadeev equation in Eq.~(\ref{eq:fadeev}) and its reduction 
to a Dirac equation. We follow the same reasoning as in Ref.~\cite{Buck:1992wz}, but with a different notation.

  We start by considering Eq.~(\ref{eq:fadeev}),
\begin{widetext}
\be \label{eq:Dirac2}
\left. \left[ g^{\alpha \beta} \delta_{j'k'} \delta^{\bar{j}' \bar{k}'} \delta_{jk} \delta^{\bar{j} \bar{k}} - \int \frac{d^4 k}{(2\pi)^4} 
L_{j k}^{\bar{j} \bar{k}, \alpha \beta}(P^2,q,k) \right]  X_{k}^{\bar{k} \beta} (P^2,q) \ \right|_{P^2=M_B^2} = 0 \ , \ee
\end{widetext}
where its kernel was defined in Eq.~(\ref{eq:kernel}). Notice that we have denoted the color terms of the equation 
by primed indices, to distinguish them from flavor factors.

For convenience, we define a new baryon wave function $Y$ by integrating over the momentum $q$~\cite{Buck:1992wz},
\be  Y_{k}^{\bar{k} \beta} (P^2) \equiv \int \frac{d^4 q}{(2\pi)^4}  X_{k}^{\bar{k} \beta} (P^2,q) \ . \ee
To express the Fadeev equation in terms of the new wavefunction one integrates Eq.~(\ref{eq:Dirac2}) over $q$ to get
\begin{widetext}
\ba & &\left.
\left[ g^{\alpha \beta}  \delta_{j'k'} \delta^{\bar{j}' \bar{k}'} \delta_{jk} \delta^{\bar{j} \bar{k}} Y_{k}^{\bar{k} \beta} (P^2) - \int \frac{d^4 q}{(2\pi)^4} \int \frac{d^4 k}{(2\pi)^4} 
 T^{\bar{k}'}_{j'l'} \tau^{\bar{k}}_{jl} \  \Gamma^{\gamma} \  S_l (-q-k) \ \ T^{\bar{j}'}_{l'k'}  \tau^{\bar{j}}_{lk} \right.\right. \nn \\
 &\times & \left.\left.
\Gamma^{\alpha} \ \
S_k (P/2+q) \ it^{\gamma \beta}_{\bar{k}} (P/2-q)
\ \ X_{k}^{\bar{k} \beta} (P^2,k)
 \right]   \ \right|_{P^2=m_B^2}  =0 \nn \ , \ea
\end{widetext}
where we have substituted the kernel given in Eq.~(\ref{eq:kernel}). Notice that it is not possible to express the equation only in terms of $Y_{k}^{\bar{k} \beta} (P^2)$ because 
there is a quark propagator that depends on the momentum $k$. Therefore, this equation is nonseparable and can be only
solved with numerical techniques~\cite{Ishii:1995bu}.

  In the so-called ``static approximation''~\cite{Buck:1992wz} one neglects the momentum dependence of the quark propagator
by assuming that the dress quark mass is much larger than the typical $k$:
\be  S_l (-q-k)=\frac{1}{ -\slashed{q}-\slashed{k} - m_l} \rightarrow \frac{-1}{m_l} \unit \ . \ee

  This approximation makes the Fadeev equation separable, and allows for a trivial integration on $k$.
The equation is reduced to the simpler form:

\ba &  & \label{eq:toproject}
\left. \left[ g^{\alpha \beta} \delta_{j'k'} \delta^{\bar{j}' \bar{k}'} \delta_{jk} \delta^{\bar{j} \bar{k}}   +  \int \frac{d^4 q}{(2\pi)^4} 
 T^{\bar{k}'}_{j'l'} T^{\bar{j}'}_{l'k'}  \tau^{\bar{k}}_{jl} \tau^{\bar{j}}_{lk} \  \Gamma^{\mu} \Gamma_{\mu} \  \frac{1}{m_l} \  \right. \right. \nn \\
  &\times & \left. \left. S_k \left(\frac{P}{2}+q \right) \ it^{\alpha \beta}_{\bar{k}} \left( \frac{P}{2}-q \right)
  \right]   Y_{k}^{\bar{k} \beta} (P^2)  \right|_{P^2=m_B^2} =0  \nn \ . \ea

 To obtain the baryon masses of the different states, we project this equation onto physical states $B$ and $B'$. The
flavor projectors are defined in App.~\ref{app:projections} for those states belonging to the octet and the decuplet
representations. Applying these projectors we find
\be  \delta_{jk} \delta^{\bar{j} \bar{k}} \ {\cal P}^{B,\dag}_{j \bar{j}}  {\cal P}^{B'}_{\bar{k} k} =
{\cal P}^{B,\dag}_{j \bar{j}}  {\cal P}^{B'}_{\bar{j} j} = \delta^{BB'} \ . \ee
In color space we take the projector onto the singlet state, ${\cal P}^{color}_{{\bar j}'j'}=\delta_{{\bar j}'j'}/\sqrt{3}$.
In the first term of Eq.~(\ref{eq:toproject}) one has
\be {\cal P}^{color,\dag}_{j'{\bar j}'} \ \delta_{j'k'} \delta^{\bar{j}' \bar{k}'} \ {\cal P}^{color,\dag}_{{\bar k}'k'}
= \frac{1}{3} \textrm{ tr}_c \ \unit =1 ,   \ee
whereas in the second term of Eq.~(\ref{eq:toproject})
\be {\cal P}^{color,\dag}_{j'{\bar j}'} T^{\bar{k}'}_{j'l'} T^{\bar{j}'}_{l'k'}  {\cal P}^{color,\dag}_{{\bar k}'k'}
= \frac{1}{3}  T^{k'}_{j'l'} T^{j'}_{l'k'}  = -2 ,   \ee
where we have used Eq.~(\ref{eq:color3bar}).

To simplify the notation we can define the matrix $M^{BB'}$

\begin{widetext}
\be \label{eq:diracM} M^{BB',\alpha \beta} (P) \equiv \frac{2}{m_l} \int \frac{d^4 q}{(2\pi)^4} \
 {\cal P}^{\dag,B}_{j \bar{j}} {\cal P}^{B'}_{\bar{k} k} \ \ 
 \tau^{\bar{k}}_{jl} \ \tau^{\bar{j}}_{lk} \ \  \Gamma^{\mu} \Gamma_{\mu}
 \ \ S_k \left( \frac{P}{2}+q \right) \ it^{\alpha \beta}_{\bar{k}} \left( \frac{P}{2}-q \right)  \ . \ee
\end{widetext}

  The final equation is expressed as a Dirac-like equation,
\be \label{eq:finalDirac} g^{\alpha \beta} \delta^{BB'} - M^{BB',\alpha \beta} (P^2=M_B^2)  = 0 \ , \ee
which is a matrix equation in Dirac and flavor spaces. 

For the baryon octet (composed by scalar diquarks) the $\alpha,\beta$-Lorentz indices are absent
and $\Gamma^\mu = i\gamma^5$. Therefore one has (after performing the change of variables $q \rightarrow -q+P/2$)
\begin{widetext}
\be \label{eq:Moctetpre} M^{BB'} (P) = - \frac{2}{m_l}  \
 {\cal P}^{\dag,B}_{j \bar{j}} {\cal P}^{B'}_{\bar{k} k} \ 
 \tau^{\bar{k}}_{jl} \ \tau^{\bar{j}}_{lk} \ 
\int \frac{d^4 q}{(2\pi)^4} \ S_k \left(P-q \right) \ it_{\bar{k}} \left( q \right)  \ . \ee
\end{widetext}
where the scalar diquark propagator $t_{\bar{k}}$ is taken from Eq.~(\ref{eq:propscalar}). 

In flavor space this equation is diagonal except for the
$\Lambda-\Sigma^0-\unit$ mixing, where one needs to solve
\be \det \left( 
\begin{array}{ccc}
1-M^{\Sigma^0 \Sigma^0} & -M^{\Sigma^0 \Lambda} & -M^{\Sigma^0\unit} \\
-M^{ \Lambda \Sigma^0} & 1-M^{ \Lambda \Lambda} & -M^{ \Lambda \unit} \\
-M^{\unit\Sigma^0} & -M^{\unit  \Lambda} & 1-M^{\unit\unit}  
\end{array}
\right)=0 \ . \label{eq:barsystem} \ee

In the isospin limit ($m_u = m_d$) one has $M^{\unit \Sigma^0}=M^{\Sigma^0 \unit}=M^{ \Lambda \Sigma^0}=M^{\Sigma^0  \Lambda}=0$.
Thus, the $\Sigma^0$ decouples from the system~(\ref{eq:barsystem}), and in this limit its mass is degenerate with the 
mass of the $\Sigma^+$ baryon. However, the elements $M^{\unit  \Lambda}$ and $M^{ \Lambda \unit}$ are nonzero and this produces a mixing
between the flavor singlet and the $\Lambda$. Therefore, to obtain the mass of the $\Lambda$ we need to solve the two-channel equation.

For the members of the decuplet (composed by axial diquarks) one uses $\Gamma^\mu = \gamma^\mu$ to get
\begin{widetext}
\be \label{eq:Mdecupletpre} M^{BB',\alpha \beta} (P) \equiv \frac{8}{m_l}  \
 {\cal P}^{\dag,B}_{j \bar{j}} {\cal P}^{B'}_{\bar{k} k} \ \ 
 \tau^{\bar{k}}_{jl} \ \tau^{\bar{j}}_{lk} \ 
 \int \frac{d^4 q}{(2\pi)^4}  S_k \left( P-q \right) \ it^{\alpha \beta}_{\bar{k}} \left( q \right)  \ , \ee
\end{widetext}
with the diquark propagator taken from Eq.~(\ref{eq:propaxial}). In this case, the Eq.~(\ref{eq:finalDirac}) is
 diagonal in flavor space.

\section{Physical baryon projections\label{app:projections}}

  Baryon masses are computed by solving the Fadeev equation projected into the different physical states. The
baryon projectors project the general wave function onto the wave functions of specific baryons. In our scheme, the baryon wave functions are the direct product of quark and diquarks wave functions. In this appendix we provide the precise expressions for completeness~\cite{Lichtenberg:1967zz,Hanhart:1995tc}.

  In Table~\ref{tab:wavefunctions} we present the baryon octet and decuplet wave functions in terms of the 
quark-diquark states. As in the main text, scalar diquarks are represented by square brackets and axial diquarks by parentheses.
We remind that in this work we neglect the axial-diquark contribution to the members of baryon octet.
\begin{widetext}
\begin{table*}
\begin{center}
\begin{tabular}{|c|c||c|c|}
\hline
Octet member &  Wave function & Decuplet member& Wave function\\
\hline
\hline
$p$ & $| u[ud] \rangle$ & $\Delta^{++}$ &   $| u (uu) \rangle$ \\
$n$ & $|d [ud] \rangle$ & $\Delta^{+}$ &  $ \frac{1}{\sqrt{3}} ( | d (uu) \rangle + \sqrt{2} | u (ud) \rangle )$  \\
$\Lambda$ & $\frac{1}{\sqrt{6}} ( |u [ds] \rangle + |d[us] \rangle - 2 |s [ud] \rangle) $ &
 $\Delta^{0}$ & $ \frac{1}{\sqrt{3}} ( \sqrt{2} | d (ud) \rangle + | u (dd)\rangle )$  \\ 
$\Sigma^+$ & $| u[us] \rangle$ & $\Delta^{-}$ &  $ | d (dd) \rangle $  \\ 
$\Sigma^0$ & $ \frac{1}{\sqrt{2}} (| u[ds] \rangle - |d[us] \rangle )	 $ & $\Sigma^{*}$ &  
$ \frac{1}{\sqrt{3}} ( | s (uu) \rangle + \sqrt{2} | u (us)\rangle )$  \\ 
$\Sigma^-$ & $|d[ds] \rangle$ & $\Sigma^{*0}$ &  $ \frac{1}{\sqrt{3}} ( | s (ud) \rangle + 
 | d (us)\rangle + | u (ds) \rangle )$  \\ 
$\Xi^0$ & $|s [us] \rangle$ & $\Sigma^{*-}$ & $ \frac{1}{\sqrt{3}} ( | s (dd) \rangle + \sqrt{2} | d (ds)\rangle )$  \\
$\Xi^-$ & $|s [ds] \rangle$ & $\Xi^{*0}$ &  $ \frac{1}{\sqrt{3}} ( \sqrt{2} | s (us) \rangle + | u (ss)\rangle )$  \\ \cline{1-2}
  Singlet  & Wavefunction  & $\Xi^{*-}$ &   $ \frac{1}{\sqrt{3}} ( \sqrt{2} | s (ds) \rangle + | d (ss)\rangle )$  \\ \cline{1-2}
$\unit$ & $ \frac{1}{\sqrt{3}} ( | u [ds] \rangle + | d [us] \rangle + | s [ud] \rangle ) $ & $\Omega^{-}$ & $ | s(ss) \rangle$ \\
\hline
\end{tabular}
\caption{\label{tab:wavefunctions} Baryon wave functions for all the members of the octet and decuplet flavor representations.}
\end{center}
\end{table*} 
\end{widetext}

These wave functions help us to construct the baryon projectors. They satisfy
\be \label{eq:hermitian} ({\cal P}_{\bar{i}j}^{B})^\dag = {\cal P}_{j\bar{i}}^{B} \ , \ee
where $B$ represents the physical baryon state. They are orthonormal within the same representation,
\be \label{eq:orthonormal} ({\cal P}_{j\bar{i}}^{B})^\dag  {\cal P}_{\bar{i} j}^{B'} = \delta^{BB'} \ . \ee

For the baryon octet they read
\ba {\cal P}_{\bar{i}j}^{p} &=& \frac{1}{2} \left( \lambda^4 - i \lambda^5 \right)_{\bar{i}j} \ , \\
{\cal P}_{\bar{i}j}^{n} &=& \frac{1}{2} \left( \lambda^6 - i \lambda^7 \right)_{\bar{i}j} \ , \\
{\cal P}_{\bar{i}j}^{\Lambda} &=& {\cal P}_{\bar{i}j}^8 = \sqrt{\frac{1}{2}} \lambda^8_{\bar{i}j} \ , \\
{\cal P}_{\bar{i}j}^{\Sigma^0} &= & {\cal P}_{\bar{i}j}^3 =  \sqrt{ \frac{1}{2}} \lambda^3_{\bar{i}j} \ , \\
{\cal P}_{\bar{i}j}^{\Sigma^{\pm}} &=& \frac{1}{2} \left( \lambda^1 \mp i \lambda^2 \right)_{\bar{i}j} \ , \\
{\cal P}_{\bar{i}j}^{\Xi^0} &=& \frac{1}{2} \left( \lambda^6 + i \lambda^7 \right)_{\bar{i}j} \ ,\\
{\cal P}_{\bar{i}j}^{\Xi^-} &=& \frac{1}{2} \left( \lambda^4 + i \lambda^5 \right)_{\bar{i}j} \ . \ea
Finally, we need the projector for the flavor singlet state, 
\be {\cal P}_{\bar{i}j}^{\unit} = \sqrt{ \frac{1}{3} } \ \unit_{\bar{i}j} \ . \ee
Note that for the baryon octet, the diquark index $\bar{i}$ runs from 1 to 3 because the scalar diquark belongs to
the antitriplet representation of $SU_f(3) \times SU_f(3)$. In particular, the $[ds],[us]$ and $[ud]$ diquarks 
are represented by $\bar{i}=1,2,3$, respectively.

For the baryon decuplet the projectors read
\ba {\cal P}_{\bar{i}j}^{\Delta^{++}} &=& \delta_{\bar{i}1} \delta_{j1}  \ , \\
{\cal P}_{\bar{i}j}^{\Delta^{+}} &=& \frac{1}{\sqrt{3}}  \delta_{\bar{i}1} \delta_{j2}+ \sqrt{\frac{2}{3}}  \delta_{\bar{i}2} \delta_{j1}  \ , \\
{\cal P}_{\bar{i}j}^{\Delta^{0}} &=& \sqrt{\frac{2}{3}} \delta_{\bar{i}2} \delta_{j2} + \frac{1}{\sqrt{3}}  \delta_{\bar{i}3} \delta_{j1}  \ , \\
{\cal P}_{\bar{i}j}^{\Delta^{-}} &=&  \delta_{\bar{i}3} \delta_{j2}  \ , \\
{\cal P}_{\bar{i}j}^{\Sigma^{*+}} &=&  \frac{1}{\sqrt{3}}  \delta_{\bar{i}1} \delta_{j3} + \sqrt{\frac{2}{3}} \delta_{\bar{i}4} \delta_{j1}  \ , \\
{\cal P}_{\bar{i}j}^{\Sigma^{*0}} &=&  \frac{1}{\sqrt{3}} (  \delta_{\bar{i}2} \delta_{j3}+ \delta_{\bar{i}4} \delta_{j2}  +  \delta_{\bar{i}5} \delta_{j1})  \ ,  \\
{\cal P}_{\bar{i}j}^{\Sigma^{*-}} &=& \frac{1}{\sqrt{3}} \delta_{\bar{i}3} \delta_{j3} + \sqrt{\frac{2}{3}}  \delta_{\bar{i}5} \delta_{j2}  \ , \\
{\cal P}_{\bar{i}j}^{\Xi^{*0}} &=& \sqrt{\frac{2}{3}}  \delta_{\bar{i}4} \delta_{j3} + \frac{1}{\sqrt{3}}  \delta_{\bar{i}6} \delta_{j1}  \ , \ea
\ba {\cal P}_{\bar{i}j}^{\Xi^{*-}} &=& \sqrt{\frac{2}{3}} \delta_{\bar{i}5} \delta_{j3}  + \frac{1}{\sqrt{3}} \delta_{\bar{i}6} \delta_{j2} \ , \\
 {\cal P}_{\bar{i}j}^{\Omega^-} &=& \delta_{\bar{i}6} \delta_{j3} \ . \ea

In this sector, the axial diquarks belong to the sextet representation of $SU_f(3) \times SU_f(3)$. Therefore, 
the index $\bar{i}$ runs from $1,...,6$ representing the diquarks $(uu),(ud),(dd),(us),(ds)$ and $(ss)$, respectively.

All the projectors satisfy explicitly Eqs.~(\ref{eq:hermitian}) and (\ref{eq:orthonormal}). \\

\section{Quark-Diquark polarization function~\label{app:barpol}}

  We will detail here the reduction of the quark--scalar diquark polarization function in Eq.~(\ref{eq:qdiqpol}). At finite temperature, the polarization function (\ref{eq:qdiqpol}) reads
\be \Pi_{k \bar{k}} (i\nu_l,{\bf p}) = T \sum_n \int \frac{d^3q}{(2\pi)^3} S_k (i\nu_l-i\omega_n,{\bf P-q}) \ t_{\bar{k}} (i\omega_n,{\bf q})  \ , \ee 
where $i\nu_l$ is a fermionic Matsubara frequency which will be analytically continued to real values at the end of the calculation.
The variable $i\omega_n$ is a bosonic Matsubara frequency appearing in the diquark propagator in the pole approximation~(\ref{eq:propscalar}),
\begin{widetext}
\be \Pi_{k \bar{k}} (i\nu_l,{\bf P}) = - T g^2_{[qq]\rightarrow qq} \sum_n 
\int \frac{d^3q}{(2\pi)^3} 
\frac{1}{(i \omega_n + \epsilon_{\bar{k},q} ) (i\omega_n - \epsilon_{\bar{k},q})} \ \frac{(P_0 - i\omega_n  ) \gamma_0 + m_k \unit}{ (i \omega_n - P_0 +E_{k,q}) 
(i \omega_n -P_0-E_{k,q}) } \ , \ee
\end{widetext}
with $\epsilon^2_{\bar{k},q}=m_{DIQ,\bar{k}}^2+{\bf q}^2 $ and $E^2_{k,q}=m_k^2+( {\bf P-q})^2$ (note that $k$ represents the quark flavor, not a momentum).

The Matsubara summation is performed taking into account the four poles using standard techniques~\cite{Kapusta:2006pm}.
We can express the final result in terms of four functions (one coming from each pole):
\be \Pi_{k \bar{k}} (P_0,{\bf P}=0) = -\frac{g^2_{[qq]\rightarrow qq}}{8\pi^2} \left( J^F_{+} +J^F_{-} +J^B_{+} +J^B_{-}  \right) \ , \ee
where we have considered the baryon at rest ${\bf P}=0$ and performed the analytical continuation $i\nu_l \rightarrow P_0+i\epsilon$.
We have also defined four $J$ functions, whose real and imaginary parts are given by
 \begin{widetext}
\ba
 \Re J^F_\pm & = & \frac{1}{2P_0} \textrm{P.V.} \int_{m_k}^{\Lambda_E} dE_{k,q}  \left[ 1- 2 n_F ( \pm E_{k,q}) \right] 
\ (\unit m_k \mp \gamma_0 E_{k,q}) \frac{ \sqrt{E_{k,q}^2-m_k^2}}{ E_{k,q} - E_\pm} \ , \\
\Im J^F_\pm &=&   \frac{\pi}{2P_0} (\unit m_k \mp \gamma_0 E_\pm) \sqrt{(E_\pm)^2-m_k^2}
\left[ 1- 2 n_F (\pm E_\pm) \right]  \Theta(\Lambda_E - E_\pm) \Theta ( E_\pm-m_k) \ , \\
\Re J^B_\pm &=& \frac{1}{2P_0} \textrm{P.V.} \int_{m_{DIQ,\bar{k}}}^{\Lambda_{E^*}} d\epsilon_k \left[ 1+ 2 n_B
 (\epsilon_k) \right]
 \ [\mp \unit m_k + \gamma_0 (\epsilon_k \mp P_0)] \frac{ \sqrt{\epsilon_k^2-m_{DIQ,\bar{k}}^2}}{ \epsilon_k - \epsilon_\pm} \ , \\
\Im J^B_\pm &=& \frac{\pi}{2P_0} [\mp \unit m_k + \gamma_0 (\epsilon_\pm \mp P_0)] \sqrt{( \epsilon_\pm)^2-m_{DIQ,\bar{k}}^2}
\left[ 1+ 2 n_B (\epsilon_\pm) \right]  \Theta(\Lambda_{E^*} -\epsilon_\pm) \Theta ( \epsilon_\pm-m_{DIQ,\bar{k}}) \ , 
\ea
\end{widetext}
with $\Lambda_E=\sqrt{\Lambda^2+m_k^2}, \Lambda_{E^*}=\sqrt{\Lambda^2+m_{DIQ,\bar{k}}^2}$. In addition,
\ba E_\pm &=& \pm \frac{m_{DIQ,\bar{k}}^2-m_k^2-P_0^2}{2 P_0} \ , \\
 \epsilon_\pm &=& \mp \frac{m_{DIQ,\bar{k}}^2-m_k^2+P_0^2}{2 P_0} \ , 
\ea
and the Fermi and Bose functions $n_F (E_k) = \left( e^{E_k/T}+1 \right)^{-1}, n_B (\epsilon_k)= \left( e^{\epsilon_k/T}-1\right)^{-1}$.

The axial diquark case ---that we have omitted for simplicity--- is straightforward. This case differs in the
explicit appearance of Dirac indices in Eq.~(\ref{eq:qdiqpol}). In addition, note that Eq.~(\ref{eq:qdiqpol})
carries an opposite sign to Eq.~(\ref{eq:q-diq}), but this cancels with the different sign in the diquark propagators
[cf. Eqs.~(\ref{eq:propscalar}] and (\ref{eq:propaxial})). We have neglected the term proportional to $p^\mu p^\nu$
in Eq.~(\ref{eq:propaxial}), which is suppressed by the diquark mass squared.

\end{document}